\documentclass[journal]{IEEEtran}
\pdfoutput=1
\usepackage{graphicx}
\usepackage{amsmath}
\usepackage{amssymb}
\usepackage[noadjust]{cite}
\usepackage{color}
\usepackage{url}
\usepackage{dblfloatfix} 
\usepackage{graphicx}
\usepackage{xfrac}
\usepackage{amsfonts}
\usepackage{float}
\usepackage{color}
\usepackage{tablefootnote}
\usepackage[flushleft]{threeparttable}
\usepackage{cite}
\usepackage{epstopdf}
\usepackage{bm}
\usepackage{subcaption}
\usepackage{relsize}
\usepackage{mathtools}
\usepackage{array}
\newcolumntype{P}[1]{>{\centering\arraybackslash}p{#1}}
\usepackage{array}
\usepackage{multirow}
\usepackage{multicol}
\usepackage{mathtools}

\DeclarePairedDelimiter{\ceil}{\lceil}{\rceil}
\usepackage{xcolor}
\usepackage{amssymb}
\usepackage{pifont}
\newcommand{\cmark}{\ding{51}}%
\newcommand{\xmark}{\ding{55}}%
\newcolumntype{P}[1]{>{\centering\arraybackslash}p{#1}}

\graphicspath{ {figures/} } 
\begin{document}
\title{Base Station and Passive Reflectors Placement for Urban mmWave Networks
} 
\author{Chethan Kumar Anjinappa,
Fatih Erden, and \.{I}smail G\"{u}ven\c{c},~\IEEEmembership{Senior Member, IEEE} 
\thanks{C.K. Anjinappa, F. Erden, and \.{I}smail G\"{u}ven\c{c} are with the Department of Electrical and Computer Engineering, North Carolina State University, Raleigh, NC 27606 (e-mail:~\{canjina,ferden,iguvenc\}@ncsu.edu).}
\thanks{This work has been supported in part by NSF through CNS-1916766 and CNS-1618692.}
\vspace{-.5cm}
}

\renewcommand{\baselinestretch}{.965}

\maketitle

\begin{abstract}
The use of millimeter-wave (mmWave) bands in 5G networks introduces a new set of challenges to network planning. Vulnerability to blockages and high path loss at mmWave frequencies require careful planning of the network to achieve a desired service quality. In this paper, we propose a novel 3D geometry-based framework for deploying mmWave base stations (gNBs) in urban environments by considering first-order reflection effects. We also provide a solution for the optimum deployment of passive metallic reflectors~(PMRs) to extend radio coverage to non-line-of-sight~(NLoS) areas. In particular, we perform visibility analysis to find the direct and indirect visibility regions, and using these, we derive a \textit{geometry-and-blockage-aided path loss model}. We then formulate the network planning problem as two independent optimization problems, placement of gNB(s) and PMRs, to maximize the coverage area with a certain quality-of-service constraint and minimum cost. We test the efficacy of our proposed approach using a generic map and compare our simulation results with the ray tracing solution. Our simulation results show that considering the first-order reflections in planning the mmWave network helps reduce the number of PMRs required to cover the NLoS area and the gNB placement aided with PMRs require fewer gNBs to cover the same area, which in turn reduces the deployment cost.\vspace{-.15cm}
\end{abstract}

\begin{IEEEkeywords}
5G, coverage enhancement, gNB location optimization, maximum coverage,  metallic reflectors, mmWave.\vspace{-.1cm}
\end{IEEEkeywords}

\section{Introduction}
\IEEEPARstart{F}{ifth}-generation~(5G) networks are expected to use millimeter-wave~(mmWave) bands to deliver high data rates with low latency and high spectral efficiency, exceeding what is possible with the traditional sub-6 GHz cellular systems~\cite{rangan2014millimeter,roh2014millimeter}. However, due to their higher frequencies, mmWave signals experience higher path loss and are more sensitive to blockages than sub-6~GHz signals, creating \emph{shadowed} regions~\cite{samuylov2016characterizing} in urban scenarios. These problems make the coverage of mmWave networks highly dependent on the link state (i.e., line-of-sight~(LoS) or non-LoS~(NLoS)) of a spatial location, and hence the layout/geometry of the environment.
To compensate for severe signal attenuation, especially in NLoS scenarios, mmWave systems use large antenna arrays with high-gain and narrow beams, which can be particularly helpful for locations that can be reached after reflections from the surrounding scatterers. Due to these unique characteristics of the mmWave signals and the special solutions required to overcome the problems associated with them, the traditional network planning approaches used at sub-6 GHz bands are not suitable at mmWave bands.

{
Most of the existing studies in the literature on mmWave network planning ignore the link outage due to frequent blockages by obstacles. More importantly, they overlook the NLoS effects, i.e., the effect of significant first-order reflections. Ignoring the NLoS effects while planning the network may result in an overestimated number of 5G mmWave base stations~(gNBs) to meet the minimum quality-of-service~(QoS) constraints, especially in NLoS areas. This, in turn, can substantially increase the gNB deployment cost as well as power consumption, and also the frequency of inter-cell handover events. For a more accurate coverage analysis based on a given set of gNB locations, one can exploit the NLoS effects and the link outage effects due to mobile obstacles, e.g. by using appropriate 3GPP blockage models~\cite{3gpp2017study} in the path loss calculation. These considerations for accurate coverage modeling may in turn reduce the number of required mmWave gNBs to be deployed.  

In addition to deploying mmWave gNBs, use of passive/active repeaters can further enhance the coverage performance of mmWave networks. Recently, it is shown that an economical means of enhancing the mmWave signals coverage for NLoS scenarios can be achieved by deploying passive metallic reflectors (PMRs)~\cite{khawaja2019effect,khawaja2018coverage,khawaja2020coverage}. Such reflectors may already be available in the environment (e.g., traffic signs, metallic billboards, light poles), or they may be placed strategically by a network operator. The deployment of PMRs in a pilot outdoor scenario is reported to enhance the coverage by a median gain of $20$ dB for NLoS locations~\cite{khawaja2019effect}. Thus, deploying PMRs along with the gNBs can be a good choice for extensively improving the coverage area in urban environments in an economical way. 
Keeping these in mind, our main contributions can be summarized as follows.} \looseness = -1

\begin{table*}[t!]
\centering
\caption{Literature review of mmWave network planning and gNB/PMR deployment.}
\label{LITERATURE_REVIEW}
  \begin{threeparttable}
\begin{tabular}{p{1.15cm} p{7.25cm} p{1.5cm} P{1.cm} P{1.75cm} P{1.2cm} P{1.45cm} }
\hline
Ref. & Goal & Scenario & 3D Planning & {\scriptsize Reflection $\&$ Blockage Effects} & PMR Deployment  &  Technique  \\
\hline
\cite{palizban2017automation,szyszkowicz2016automated} & Maximize LoS coverage area & Outdoor-U & \xmark & \xmark & \xmark & Relaxed IP \\
\hline 
\cite{mavromatis2019efficient} & Maximize LoS coverage area achieving minimum QoS
 & Outdoor-U & \xmark & \xmark & \xmark &  IP  \\
\hline
\cite{dong2019cost} & Connectivity-constrained minimum
cost BS deployment & Outdoor-UM & \xmark & \xmark & \xmark & IP \\
\hline
\cite{soorki2017joint} & Minimize number of BSs achieving minimum LoS coverage & Outdoor & \cmark & \xmark & \xmark & CCS-IP \\
\hline
\cite{fatih2020MABP} & Maximize LoS coverage using multi-armed bandit learning & Outdoor-U & \cmark & \xmark &  \xmark & MABL\\
\hline
\cite{wang2019novel} & Minimize number of BSs achieving both coverage and capacity QoS constraints & Indoor & \cmark & \xmark & \xmark & IP \\
\hline
\cite{danford2017end} & Minimize total cost of operation achieving reliability constraints  & Outdoor-U & \cmark & \xmark & \xmark & IP \\
\hline\cite{peng2015effective} & Numerical experiment to enhance coverage area to NLoS region (No optimization)  & Outdoor & \cmark & \xmark & \cmark & - \\
\hline
{This work} & Maximize coverage area considering gNBs and PMRs  & Outdoor-U & \cmark & \cmark & \cmark & IP \\
\hline
\end{tabular}
\begin{tablenotes}
    \item \scriptsize \textbf{Short-hand notation:} Urban (U), Urban Manhattan (UM), Integer Programming (IP), Multi-Armed Bandit Learning (MABL), and Chance-Constrained Stochastic (CCS).
  \end{tablenotes}
 \end{threeparttable}
 \label{Table:Literature}\vspace{-4mm}
\end{table*}

    \noindent $1)$ \textbf{Geometry-and-blockage-aided path loss model:} We propose a novel way of calculating the total path loss by taking into account the blockage effects and the geometry of the environment through visibility analysis. Our idea is similar in nature to~\cite{palizban2017automation,szyszkowicz2016automated,mavromatis2019efficient} that employs visibility analysis as a part of the path loss calculation. However, an important difference is that in our case we take into account the blockage effects and the reflection effects for the NLoS area. This way, besides the LoS path, we also consider the possibility of establishing a link through a first-order reflected path. We do so by finding the direct visibility and indirect visibility (reflection dominated) regions by considering two types of reflection: \textit{specular} and \textit{diffuse}. {With this, we derive the \textit{geometry-and-blockage-aided path loss model (GB-PLM)} to calculate the total path loss.}
    
    
    \noindent $2)$ \textbf{Optimizing mmWave gNB(s) and PMR(s) placement:} In contrast to existing works~\cite{palizban2017automation,szyszkowicz2016automated,mavromatis2019efficient,dong2019cost,soorki2017joint,fatih2020MABP,wang2019novel,danford2017end,peng2015effective}, we focus on the deployment and coverage of mmWave gNB(s) aided with PMRs provided the 3D map of the environment. We present a two-stage solution, where, at first, the gNB(s) are placed to maximize the coverage area by utilizing the GB-PLM.
    Then, PMRs are placed to extend the coverage area to parts of the environment not served by any gNB. 
    We pose and solve separate linear binary optimization problems to obtain the optimal locations and orientations of the gNB(s) and PMR(s),~respectively.
    
    \noindent $3)$ \textbf{Coverage and outage analysis:} We analyze the coverage fraction with and without the first-order reflection effects and the PMRs. We show by extensive simulations that considering the first-order reflections in gNB placement problem helps to reduce the number of PMRs required to cover the NLoS area. Also, the gNB placement aided with PMRs requires fewer gNBs to cover the same area. 
    The proposed solution is compared with the full-blown 3D deterministic ray-tracing (RT) simulator, Wireless InSite~\cite{WirelessInsite}. However, the results from the Wireless InSite can only be used as a benchmark for the static environment alone, as the RT cannot incorporate the effect of the mobile blockages to channel~models.

The rest of the paper is organized as follows. {Section~\ref{Sec:LITERATURE_REVIEW} presents the literature review followed by the system model and problem definition in Section~\ref{Sec:System Model}.} The preliminaries of direct and indirect visibility analysis, which are utilized to derive the GB-PLM, are presented in Section~\ref{Sec:Visibility_PL_Model}. This is followed by the formulation and the solution for the mmWave gNB placement problem in Section~\ref{Sec: gNB Placement}, and then the PMR placement problem along with the preliminaries are presented in Section~\ref{Sec:PMRs_Problem_Position_Orientation}. In Section~\ref{Section:Simulation}, we validate the efficacy of our proposed algorithms using a simple 3D map, and finally, we provide concluding remarks in Section~\ref{Sec:Conclusion}.

{\bf Notation}: Scalars, vectors, and sets are represented by non-bold letters (e.g., a, A), lower-case boldface (e.g., \textbf{a}), and calligraphic (e.g., $\mathcal{A}$) letters, respectively. Unit directional vectors are represented with a hat (e.g., $\hat{\textbf{a}}$). For an integer $K$, we use the shorthand notation $[K]$ for the set of non-negative integers $\{1,2,\ldots, K\}$. The operation $\bigcup$ and~$\bigcap$ represent the union and intersection operations on a set, respectively. Finally, the sign($\cdot$), $\ceil{\cdot}$, and ``$\cdot$" operators represent the sign, ceiling, and dot product operations, respectively.

\section{Literature Review}
\label{Sec:LITERATURE_REVIEW}
Recently, there have been a number of studies on mmWave network planning~\cite{palizban2017automation,szyszkowicz2016automated,mavromatis2019efficient,dong2019cost}. In most of these studies, as also highlighted in Table~\ref{Table:Literature}, the coverage and placement of the gNBs in an outdoor area are considered by maximizing the LoS area while maintaining a minimum QoS. These studies often ignore the contribution of strong reflections to the coverage in NLoS areas and simply presume NLoS areas will be in outage. For example, the studies in~\cite{palizban2017automation} and~\cite{szyszkowicz2016automated} consider wall-mounted mmWave gNBs while~\cite{mavromatis2019efficient} uses lamp post mmWave gNBs, and they all place the gNBs such that the LoS area is maximized. A connectivity-constrained minimum-cost mmWave gNBs deployment for a Manhattan-type geometry is studied in~\cite{dong2019cost}, where the potential coverage area is restricted to only the blockage-free regions, and the effect of low-order reflections in NLoS areas are neglected. Moreover, these studies are all based on two-dimensional~(2D) maps. However, in complex environments with high-rise buildings of different heights, multiple obstructions, and gNBs located at varying heights, a more advanced visibility analysis needs to be performed to distinguish between LoS and NLoS areas.

There are only a few studies that use a 3D map setup for coverage analysis and network planning at mmWave bands. In~\cite{soorki2017joint}, where the goal is to find the minimum number of mmWave gNBs such that the coverage constraints are satisfied under stochastic user orientation, namely, head, hand, and pocket, access points~(APs) and user equipment's~(UE) are positioned in 3D space. In this study, the authors make a hard assumption that, for each UE, there exists an LoS path to at least one of the APs. Similarly, there are other 3D network planning work for outdoor coverage~\cite{fatih2020MABP}, indoor coverage~\cite{wang2019novel}, fixed backhaul and access networks (Facebook 60 GHz Terragraph)~\cite{danford2017end}, unmanned aerial vehicle-mounted BS placement~\cite{shakoor2020joint}, small cells with hybrid back-hauls~\cite{cao2018cost}, and networking infrastructure with machine-type communication~\cite{xu2016deployment}. \looseness = -1

\begin{table*}[b!]
\vspace{-.1cm}
\renewcommand{\arraystretch}{1.15}
\centering
\caption{ Nomenclature for the commonly used notation in this paper.}
\label{Table:Notation}
  \begin{threeparttable}
\begin{tabular}{| p{.9cm} p{4.35cm} | p{.9cm} p{4.15cm}| p{.9cm} p{4.35cm}|}
\hline
Notation & Description & Notation & Description & Notation & Description  \\
\hline
$\mathcal{L}^\text{gNB}$ & Set of potential gNB 3D locations & $N$ & Number of potential gNB locations & $i$ & Index variable for $\mathcal{L}^\text{gNB}$ \\
\hline
$\mathcal{L}^\text{SA
}$ & Set of service area 3D locations & $M$ & Number of grids in the service area & $j,l$ & Index variable for $\mathcal{L}^\text{SA}, \mathcal{L}^\text{OSA}$ \\
\hline
$\mathcal{L}^\text{Build}$ & Set of 3D building locations & $B$ & Number of PMR building locations & $b$ & Index variable for $\mathcal{L}^\text{Build}$ \\
\hline
$\mathcal{L}^\text{PMR}$ & Set of PMR 3D locations & $K$ & Number of feasible PMR locations & $k$ & Index variable for $\mathcal{L}^\text{PMR}$ \\ \hline
$\mathcal{L}^{\text{gNB$\star$}}$ & Set of optimum $N^\text{gNB}$ gNB locations  & $N^\text{gNB}$ & Number of gNBs & $N^\text{PMR}$ & Number of PMRs\\
\hline
$\mathcal{L}^{\text{VgNB$\star$}}$ & Set of locations visible from $\mathcal{L}^\text{gNB$\star$}$ &  $d^\text{3D}$ & 3D Euclidean distance & $d^\text{2D}$ & 2D Euclidean distance   \\
\hline
$\mathcal{L}^\text{OSA}$ & Set of locations in outage &  $\textrm{PL}_\text{(N)LoS}$ & (N)LoS path loss & $\textrm{PL}_\text{Out}$ & Outage path loss      \\ 
\hline
$\mathcal{L}^{\text{VOSA}}$ & Set of locations visible from $\mathcal{L}^\text{OSA}$ & $\mathbb{P}_\text{(N)LoS}$ & (N)LoS probability & $\theta^\text{I(R)}$ & Incident (reflected) angle \\ \hline
$\mathcal{V}(\cdot)$ & Set of directly visible locations & $\hat{\mathbf{i}}$ & Unit-norm incident ray vector & $(\Delta) \delta $ & (In)direct indicator variables  \\ \hline
$\mathcal{V}^S(\cdot)$ & Set of specular visible locations &  $\hat{\mathbf{r}}$ & Unit-norm reflected ray vector & $\gamma_{\max}$ & MAPL threshold\\ 
\hline
 $\mathcal{V}^D(\cdot)$ & Set of diffuse visible locations & $\hat{\mathbf{n}}$ & Unit-norm normal vector & $a^{\star}$  & Area of $\star \in \{$Cell radius, SA, PMR$\}$ \\ 
\hline
\end{tabular}
 \end{threeparttable}
 \vspace{-.5cm}
\end{table*}

For NLoS communications, earlier studies, such as~\cite{Rajagopal_Reflectivity,28GHz_NYC,hosseini2020attenuation}, have shown that well-known lousy objects, such as metal, concrete, modern buildings, and the human body, act as good reflectors at mmWave bands, enabling the UE to receive reflections for NLoS communication. On the other hand, rough outdoor surfaces, such as the dressed stone wall or the bath stone wall, are seen to show large variations in the signal strength of the first-order scattered components at high frequencies (e.g., at 60~GHz~\cite{goulianos2017measurements}). Thus, the reflection profile of a scatterer is highly dependent on its material and the frequency of operation. Recently, it is shown that the mmWave signal enhancement in a NLoS environment can be achieved using PMRs~\cite{khawaja2019effect,khawaja2018coverage,khawaja2020coverage}. A PMR is used to enhance the coverage for mmWave signals at 28~GHz for NLoS propagation scenarios in an outdoor~\cite{khawaja2019effect} and indoor area~\cite{khawaja2018coverage}, respectively. An analytical model for end-to-end reflected received power for NLoS propagation is developed in~\cite{khawaja2020coverage}. In these studies, it is reported that using a 33~in~$\times$~33~in square PMR, the received power can be increased by a median of 20~dB and 19~dB in the indoor and outdoor environment, respectively, compared to a no reflector~case. 

The above use case of the PMRs is attractive due to multiple reasons. First, at mmWave frequencies, the skin depth of the metallic conductors is smaller relative to sub-6~GHz bands, and hence material penetration is low. This results in stronger reflections, which can be utilized for NLoS directional communication~\cite{ghosh2002electromagnetic}. Thus, deployment of the PMRs can create a favorable propagation environment by introducing new multi-path components~(MPCs) to the channel and increasing the overall spatial diversity of the MPCs~\cite{yang2018sense}.
Second, PMRs are low cost and low maintenance, and easy to implement. They require no power, have a higher life span, and can be used in real environments like advertisement boards, metallic art signs, lamp posts, among others. Finally, unlike intelligent reflecting surfaces (IRSs), PMRs do not require special design. Note that the PMRs are similar to the IRSs except that the IRSs are designed to provide \textit{anomalous} reflections to the desired location by carefully designing the scattered field~\cite{cao2019intelligent,ozdogan2019intelligent,basar2019wireless}.

From the above perspective, it is clear why deploying PMRs along with the gNBs might be necessary for optimal-economical improvement of the coverage area. This is illustrated in~\cite{peng2015effective} for an environment with a fixed gNB, where multiple large-sized PMRs are used to reflect the mmWave signals to the NLoS regions with low received power levels. However, the authors of this study do not aim to optimize the gNB and reflector locations. Instead, they only consider roof-top position for the PMRs. To the best of our knowledge, no other attempt has been made to optimize the placement of gNBs (considering the reflection effects) and PMRs simultaneously with the aim of improving the mmWave signal coverage. {To make the paper self-contained, a comparison of how this work differs from the current literature is provided in Table~\ref{Table:Literature}.}


\section{System Model and Problem Statement}\label{Sec:System Model}
In this section, we will introduce the system model as well as the mmWave gNB/PMR placement problems. {Frequently used nomenclature are presented in Table~\ref{Table:Notation}.}

\subsection{3D Environmental Model}
We assume an outdoor environment whose true 3D map, which includes terrain data, landmarks, street routes, etc., is available. Note that free 3D geographic data of some of the real environments can be obtained from publicly available OpenStreetMap (OSM)~\cite{haklay2008openstreetmap} or several applications that generate 3D buildings from OSM such as OSM-3D, OSM2World, OSM Building, etc. It is also worth noting that street-level photos can also be used to generate maps based on Mapillary service.

\begin{figure}[!t]
    \centering
    \includegraphics[scale=.275]{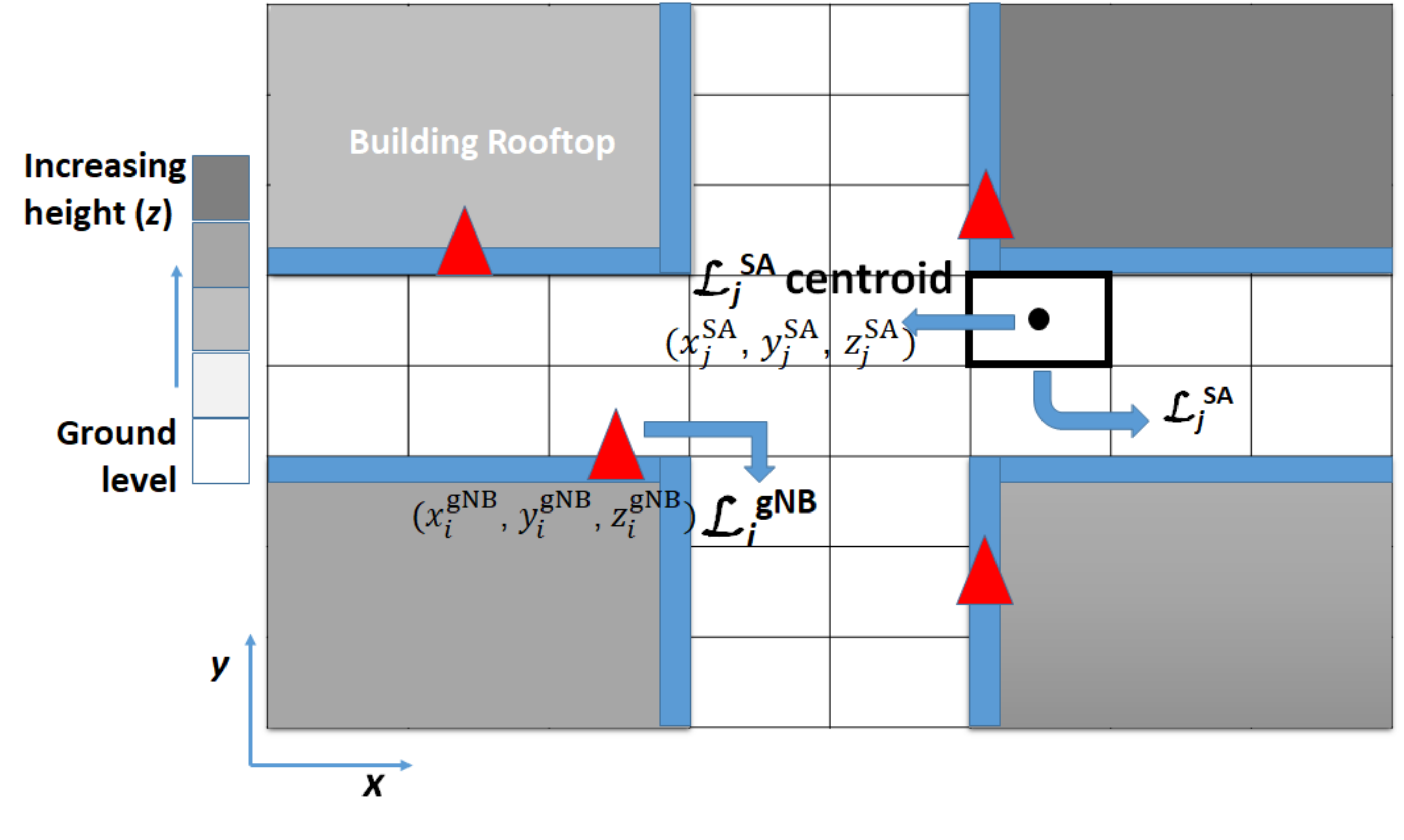}
    \caption{2D illustration of the DEM with SA and gNB on a map. The square grids and triangles indicate the SAs and possible gNB locations,~respectively.}
    \label{fig:Service_Area}
\end{figure}

Using such a 3D map, a digital elevation model~(DEM) of the map in the form of a raster (a grid of squares), which holds the 3D coordinates of the centroid of the grid (2D coordinates and the height data) can be generated. The DEMs are the most common basis for representing digital maps whose quality depends on the desired resolution, which dictates the number of grids, increasing which results in more accurate maps, but also results in higher computational complexity for the optimization algorithm. This notion of the grids will thus be used in the rest of the paper as it simplifies the mathematical tractability that follows.
The DEM of a sample outdoor environment is shown in Fig.~\ref{fig:Service_Area}. 


\subsection{Deployment and Coverage Model for mmWave gNBs}\label{Potential_gNB_Locations}
We consider an urban macro~(UMa) outdoor-to-outdoor communication scenario~\cite{3gpp2017study}, where the gNBs can be placed on the rooftop of the buildings and UEs are the outdoor ground users. Further, we define service area (SA) as the area that covers all possible UE locations (i.e., whole area minus the area covered by buildings).
We represent the discretized 3D-coordinates $\Re^3$ of the potential gNB locations by the set
\begin{equation}
    \mathcal{L}^\text{gNB} \in \{ (x^\text{gNB}_i, y^\text{gNB}_i, z^\text{gNB}_i)| \forall i \in [N] \},
\end{equation}
where $N$ is the total number of potential gNB locations. The potential gNB locations depend on the placement restrictions. To lower the computational cost of the optimization algorithm, candidate locations can be reduced using different approaches, such as viewshed analysis~\cite{fatih2020MABP}, and generalized voronoi diagram and natural neighbor analysis~\cite{szyszkowicz2016automated}. Similarly, the predefined SA for the UEs are denoted by the set 
\begin{equation}
\mathcal{L}^\text{SA} \in \{(x^\text{SA}_j, y^\text{SA}_j, z^\text{SA}_j)| \forall j \in [M]\}~,
\end{equation}
with $M$ being the total number of divided SAs. This is illustrated by a simple scenario in Fig.~\ref{fig:Service_Area}. 

The 3D Euclidean distance~($d^\text{3D}_{i,j}$) between the spatial location of ${i}^\text{th}$ potential gNB~($\mathcal{L}^\text{gNB}_i$) and $j^\text{th}$ SA~($\mathcal{L}^\text{SA}_j$) is denoted~as
\begin{equation}
d_{i,j}^\text{3D} = \sqrt{ ( x^\text{gNB}_i -  x^\text{SA}_j)^2 + (y^\text{gNB}_i - y^\text{SA}_j)^2 + (z^\text{gNB}_i - z^\text{SA}_j)^2}~, 
\end{equation}
which is essential for an accurate path loss calculation. Similarly, the 2D Euclidean distance ($d_{i,j}^\text{2D}$) between the spatial location of ${i}^\text{th}$ potential gNB and ${j}^\text{th}$ SA is denoted as 
\begin{equation}
d^\text{2D}_{i,j}~= \sqrt{ ( x^\text{gNB}_i -  x^\text{SA}_j)^2 + (y^\text{gNB}_i - y^\text{SA}_j)^2 }~,
\end{equation}
which is required for the blockage model LoS/NLoS link probability calculation that follows up in Section~\ref{Sec:Geometry_PL_Dynamic}.

\subsection{Deployment and Coverage Model for PMRs}\label{Sec:PMR_Loc}
We assume the reflectors can only be mounted on the building walls and outer surface of building rooftops with the help of mounting clamps and can be oriented in any direction. As an alternative to PMRs, passive metasurfaces reflectors may be used to provide desired reflection angles without necessarily changing the physical orientation of the reflector~\cite{ozdemir202028}. While the PMRs can be placed on lamp posts, traffic lights, etc, we do not explicitly consider such street-level locations in this work. On the other hand, our framework is readily applicable for all such scenarios.

We denote the set of candidate PMR locations on the buildings by $\mathcal{L}^\text{Build}$, which can be mathematically represented~as
\begin{eqnarray}
\label{eq:C_build}
\begin{aligned}
\mathcal{L}^\text{Build} & = \left\{ (x_b,y_b,z_b) \mathlarger{\mid} \; \forall b \in [B] \right\}.
\end{aligned}
\end{eqnarray}
The set $\mathcal{L}^\text{Build}$ contains the 3D coordinates of $B$ points on the outer surfaces of the buildings. Note that ideally this is a continuous space, which is not suitable for discrete optimization. For this purpose, we discretize this set into grids with the resolution being equal to the reflector size.
 
The earlier studies have shown that the coverage gains due to using a reflector not only relies on the position and the orientation of the PMR, but also depends on the shape and the size of it. The authors in~\cite{khawaja2018coverage,khawaja2020coverage} used a finite-sized aluminum flat-square, cylindrical, and spherical reflectors. They reported that received power from the flat-square reflectors is higher compared to the cylindrical and spherical reflectors. Thus, we consider only flat-square aluminum reflectors (with no tilt). However, we investigate different sizes of the PMR with negligible thickness (see Section~\ref{Section:Simulation}). Further, we assume that the deployed PMRs and the SAs are at a far-field distance from the deployed gNBs and PMRs, respectively.

\subsection{mmWave gNB/PMR Deployment Optimization}
The main goal of this paper is to ``\textit{maximize the coverage area of the SA by placing a minimum number of gNBs and PMRs from a set of potential gNB and PMR locations, respectively.}" In other words, the goal is to find the places where gNB(s) and PMR(s) should be deployed so that the maximum number of grids in the SA is covered. By definition, the $j^\text{th}$ grid in the SA $\mathcal{L}^\text{SA}$ is said to be covered if the path loss at the center of that grid is less than or equal to the maximum allowable path loss threshold (MAPL)~$\gamma_\text{max}$. If the total path loss exceeds $\gamma_\text{max}$, then the corresponding grid is deemed to be in an outage. \looseness = -1

A simple way to approach our goal of placing gNB(s) and PMRs is to divide the problem into two independent sub-problems: \textit{gNB placement} and \textit{PMR placement}. At first, we pose and solve an optimization problem for a fixed number of gNB(s) so that the coverage of the SA is maximized. Due to the outage, a part of the SA may be outside the coverage area of the placed gNB(s). In that case, PMRs are deployed to enhance/extend the coverage, where the optimal location and orientation of the PMRs need to be determined. Ideally, our goal of placing gNB(s) and PMRs should be approached in a joint manner, where both the location of the gNBs and locations/orientations of the PMRs should be solved jointly. However, this is not possible due to mathematical intractability owing to the difficulty in modeling the PMRs constraints into an optimization problem and computational complexity. Thus, we follow the proposed sub-optimal yet simpler-tractable approach so that it can show the concept and feasibility of the proposed method. 

Up next, we propose a novel GB-PLM which captures the path loss model both in terms of the computational geometry and the statistical nature of the link. The derived GB-PLM will be used in the subsequent sections, where we define and solve the independent {g}NB and PMR placement problems with the objective of maximizing the coverage area for a fixed number of gNB(s) and PMRs, respectively.

\section{Path Loss Models and Visibility Analysis}\label{Sec:Visibility_PL_Model}
A mmWave link is mainly characterized by the LoS path and a few significant low-order (mostly first-order) NLoS reflections~\cite{ChannelModel_Rappaport,khawaja2020multiple}. Naturally, a link can either be in the LoS, NLoS, or the outage state. The LoS state happens when there is no blockage through the shortest path between the transmitter and the receiver. This blockage can either be due to the \textit{static} elements in the environment, such as buildings, landmarks, etc., which are solely dependent on the geometry (or layout) of the environment, or it can also be due to the \textit{dynamic} elements in the environment, such as vehicles, human blockage, etc. The NLoS state occurs when there is a blockage of the LoS path and the communication happens only through reflections. Finally, the outage happens when the path loss (either due to the LoS or NLoS) is greater than the MAPL threshold denoted as~$\gamma_\text{max}$.

With the above perspective, we take a step toward addressing the GB-PLM simultaneously through the use of {visibility analysis with one reflection} and the 3GPP standard blockage and path loss models. We emphasize that the visibility analysis only captures the geometry-induced (static) blockages of the environment. On the other hand, the blockage models with link-state probabilities account for the dynamic blockages, which will be discussed~later.

We now sidestep away from the GB-PLM to discuss the visibility analysis concept, which will be pivotal in understanding our derived path loss model in Section~\ref{Sec:Geometry_PL_Dynamic}. The visibility analysis has been studied in computer graphics in relation to ray tracing, art-gallery problems, and even in wireless communications for the coverage of the mmWave gNBs as in~\cite{palizban2017automation,szyszkowicz2016automated,mavromatis2019efficient,dong2019cost}. Some of the important definitions of (direct and indirect) visibility analysis are in order.


\subsection{Direct Visibility Region} In a given map, a point $p$ is said to be directly visible from the point $s$ if there exists a line segment between $p$ and $s$ that does not intersect any obstacles, as illustrated in Fig.~\ref{fig:Visibility}. The set of all the points that are directly visible from the point $s$ is denoted by the set $\mathcal{V}(s)$ and can be computed in linear time~\cite{aronov1998visibility} for a given point. This is illustrated with the help of a simple 2D polygon both in Fig.~\ref{fig:Visibility}(a) and Fig.~\ref{fig:Visibility}(b). Intuitively, the set $\mathcal{V}(s)$ corresponds to the LoS area for which the LoS path can exist from the point $s$, this set is also known as star-convex set in math and computational geometry community. We emphasize that $\mathcal{V}(s)$ characterizes the propagation environment considering only its static nature. The LoS path can still be blocked by a dynamic element that will be captured in a stochastic manner later in Section~\ref{Sec:Geometry_PL_Dynamic}.

\subsection{Indirect Visibility Region}
We consider two different types of indirect visibility based on the reflection type: \textit{specular} (Fig.~\ref{fig:Visibility}(a)) and \textit{diffuse} (Fig.~\ref{fig:Visibility}(b)) reflection under the assumption of the first-order reflection. A theoretical justification for our restriction to the first-order reflection is that the first-order reflections capture the significant power for the NLoS communication, and we ignore the higher-order reflections owing to the high path loss and absorption loss at mmWave bands. It should be emphasized that we compute the indirect visibility only for the points that are not in the direct visibility region. In other words, the indirect visibility region corresponds to the region, which can be reached with a single reflection outside the direct visibility region.\looseness = -1

\subsubsection{Specular Visibility Region}
A point $p_s$ is said to be indirectly visible from the point $s$ after one specular reflection, if there exists a point $b$ on a scatterer (building, advertisement board, etc.) visible from both $p_s$ and $s$, such that the angle between the incident and the reflected wave from the point $b$ follows the standard law of reflection (Snell's law): the angle of incidence equals the angle of reflection~\cite{de2006reflections}. The set of all the points that are indirectly visible from the point $s$ via specular reflection are denoted by the set $\mathcal{V}^S(s)$. Intuitively, the set $\mathcal{V}^S(s)$ correspond to a reflection dominated region~\cite{haneda20165g}, as seen in Fig.~\ref{fig:Visibility}(a).\looseness = -1

Specular reflection commonly occurs from the smooth surfaces and is found to be an appropriate model for scattering from the terrain, buildings, and other electrically-large structures encountered in the analysis of wireless communications systems~\cite{ellingson2019path}. For a given map, the set $\mathcal{V}^S(s)$ can be computed in linear time using the algorithms that utilize the concept of \textit{mirror edges}~\cite{vaezi2017extend}, where the scatterers can make at least a part of a specific invisible segment visible to a given point or segment. Details can be found in~\cite{aronov1998visibility,vaezi2017extend}. 

\subsubsection{Diffuse Visibility Region}
\begin{figure}[!t]
\begin{subfigure}{.5\textwidth}
  \centering
  \includegraphics[scale = 0.35]{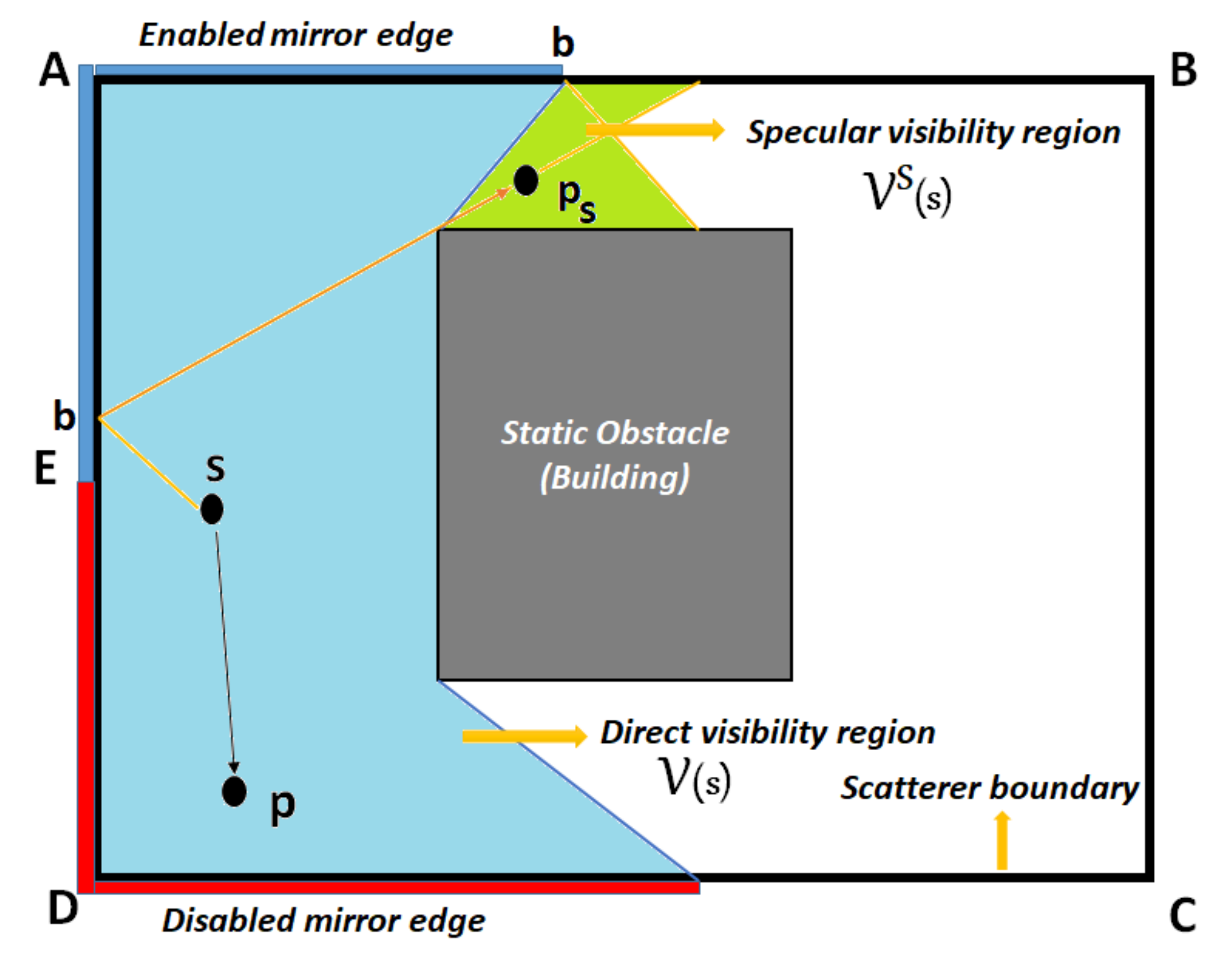}
  \vspace{-.1cm}
  \caption{ }
  \label{fig:Specular}
\end{subfigure}
\vspace{-.1cm}
\begin{subfigure}{.5\textwidth}
  \centering
  \includegraphics[scale = 0.35]{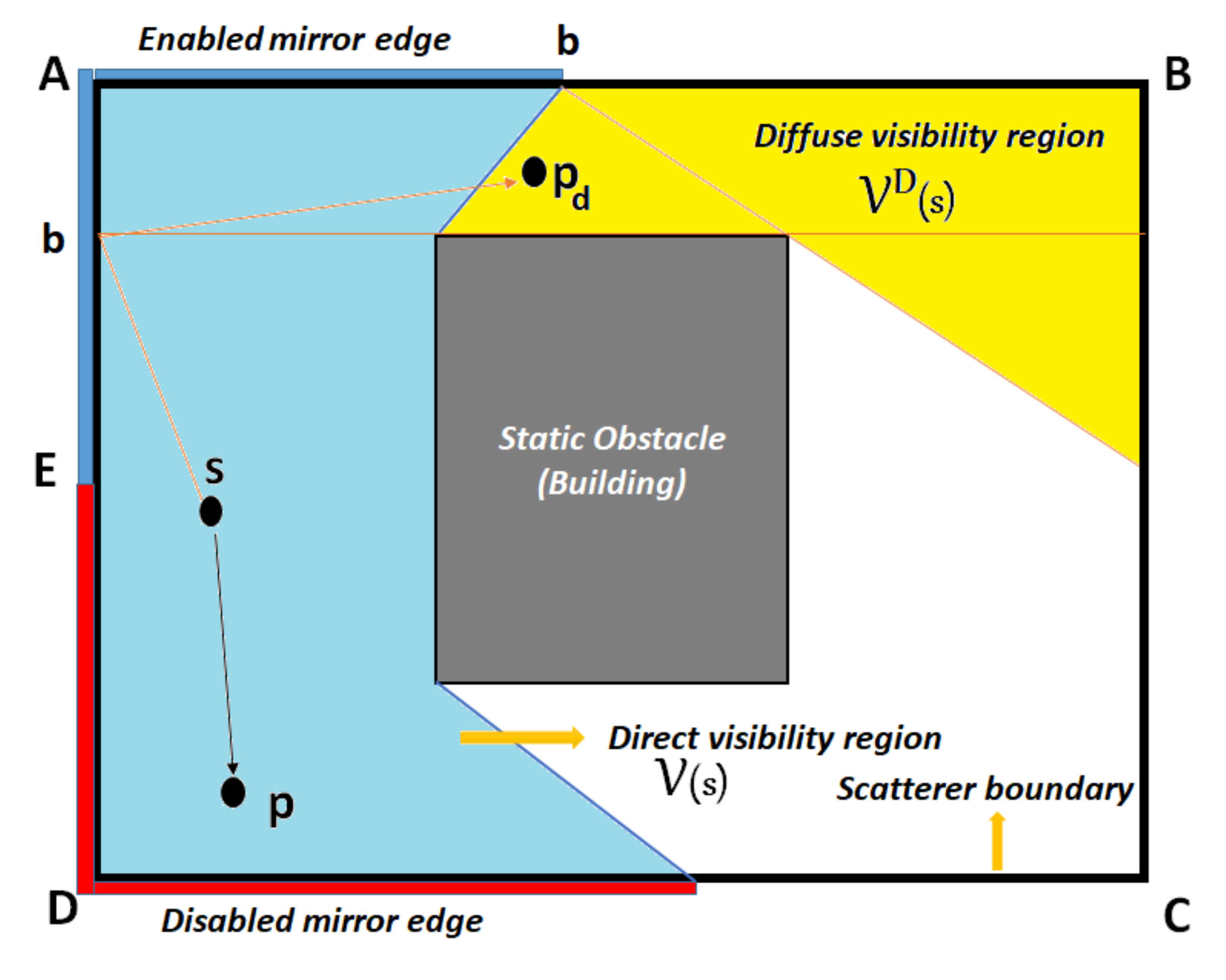}
  \vspace{-.1cm}
  \caption{ }
  \label{fig:Diffuse}
\end{subfigure}
\caption{Illustration of a 2D direct visibility and indirect visibility regions. Both the figures show the direct visibility for a fixed point of $s$. (a) The specular visibility region following Snell's law. (b) The diffuse visibility region.}
\label{fig:Visibility}
\end{figure}

A point $p_d$ is said to be indirectly visible from the point $s$ after one diffuse reflection, if there exists a point $b$ on a scatterer, visible from $p_d$ and $s$. The set of all the points that are indirectly visible from the point $s$ via diffuse reflection are denoted by the set $\mathcal{V}^D(s)$. This is illustrated in Fig.~\ref{fig:Visibility}(b). Unlike the specular visibility, the diffuse visibility calculation is not required to follow Snell's law. As noted in~\cite{solomitckii2016characterizing}, the diffuse scattering makes a noticeable contribution to the total received power for NLoS links at the mmWave bands due to the surface roughness. Thus, when considering the NLoS communication in an urban setting it is critical to include the diffuse scattering region too. Note that by construction, $\mathcal{V}^S(s) \subseteq \mathcal{V}^D(s)$.

The choice of the reflection type (specular or diffuse) depends on the scatterer material. For instance, if a scatterer material surface is smooth and flat, such as float/clear glass buildings, then the specular reflection is the correct choice of reflection. If the material has a rough surface, then diffuse scattering occurs. An example of the phenomenon of direct and indirect visibility for a fixed source point $s$ is illustrated in Fig.~\ref{fig:Visibility} for a generic 2D map. A similar approach can be adopted for the 3D maps. The black boundary points are assumed to be scatterers. The blue, green, and yellow regions correspond to the direct, specular, and diffuse visibility regions for the source at point~$s$. \looseness = -1

If the scatterer material is known, then one can incorporate it for an effective calculation of the indirect visibility region. For instance, the thick blue boundary points in Fig.~\ref{fig:Visibility} correspond to a scatterer segment (mirror edge) enabled for reflections~\cite{vaezi2017extend}, where the scatterer is assumed to be composed of materials classified as ``good reflectors" at mmWave bands such as glass and concrete. If the scatterer is of type ``bad reflectors", such as bath stone or scatterer covered with vegetation, then the mirror edge can be disabled for indirect visibility calculation. Since it is often not easy to obtain the material information, in this paper we enable all the possible mirror edges while calculating the indirect visibility. During the simulations, we considered both the specular and diffuse visibility paths. However, from here onwards, for ease of understanding the derived path loss model, we restrict our discussion of reflection effects to the specular visibility. \looseness = -1

\subsection{Geometry Aided Path Loss Model}
Based on direct/indirect visibility  of the environment alone, the {path loss $\textrm{PL}_{ij}$ between $\mathcal{L}_i^\text{gNB}$
and $\mathcal{L}_j^\text{SA}$} can be defined as

\begin{equation}\label{Equ:PL_Static}
  \textrm{PL}_{ij} = \begin{cases}  \textrm{PL}_\text{LoS}(d^\text{3D}_{i,j}),  &\mbox{if $\mathcal{L}^\text{SA}_j \in \mathcal{V}(\mathcal{L}^\text{gNB}_i)$}\\
\textrm{PL}_\text{NLoS}(d^\text{3D}_{i,j}), & \mbox{if $\mathcal{L}^\text{SA}_j \in \mathcal{V}^S(\mathcal{L}^\text{gNB}_i)$} \\
\textrm{PL}_\text{Out},& \mbox{otherwise}
\end{cases}~,
\end{equation}
which considers the LoS and NLoS connectivity regions in Fig.~\ref{fig:Visibility}.
If $\mathcal{L}^\text{SA}_j \in \mathcal{V}(\mathcal{L}^\text{gNB}_i)$, then the spatial location $\mathcal{L}^\text{SA}_j$ can have a LoS path from the location $\mathcal{L}^\text{gNB}_i$. In that case, the path loss (in dB) is based on LoS path loss model $\textrm{PL}_\text{LoS}$. If on the other hand, $\mathcal{L}^\text{SA}_j$ lies inside the indirect visibility region, then the path loss is based on NLoS path loss model $\textrm{PL}_\text{NLoS}$.
While we can use ray tracing to calculate path loss more accurately, it requires a very high computational cost and highly-detailed modeling of the propagation environment -- in this work we use 3GPP path loss models to obtain $\textrm{PL}_\text{LoS}$ and $\textrm{PL}_\text{NLoS}$~\cite{3gpp2017study}[Table 7.4.1-1: Path loss models]. If the point $\mathcal{L}^\text{SA}_j$ is not directly or indirectly visible from $\mathcal{L}^\text{gNB}_i$, then the point is in outage, as illustrated with the white regions in~Fig.~\ref{fig:Visibility}. 

The path loss calculation in~\eqref{Equ:PL_Static} leverages the geometry of the environment and hence captures the path loss better than the approaches in~\cite{palizban2017automation,szyszkowicz2016automated,mavromatis2019efficient,dong2019cost}, which categorize the service area only as LoS and NLoS. This model, however, does not include the impact of dynamic objects, such as cars, trucks, human blockages etc., which will be captured separately using the blockage models in the next sub-section. 

\subsection{Geometry and Blockage Aided Path Loss Model}\label{Sec:Geometry_PL_Dynamic}
To capture the dynamic blockage effects, we adopt the link-state probabilities from~\cite{3gpp2017study} and rewrite the path loss model in~\eqref{Equ:PL_Static} as~\cite{Hourani2016}
\begin{equation}
\label{Equ:PL_Dynamic}
\textrm{PL}_{i j}=\left\{\begin{array}{ll}
 \mathbb{P}_{\mathrm{LoS}}(d^\text{2D}_{ij}) \textrm{PL}_{\mathrm{LoS}}(d^\text{3D}_{ij})+\\  \mathbb{P}_{\mathrm{NLoS}}(d^\text{2D}_{ij}) \textrm{PL}_{\mathrm{NLoS}}(d^\text{3D}_{ij}), &  \text {if } \mathcal{L}^\text{SA}_j \in \mathcal{V}\left(\mathcal{L}^\text{gNB}_{i}\right) \\
\textrm{PL}_{\mathrm{NLoS}}(d^\text{3D}_{ij}), &\text {if } \mathcal{L}^\text{SA}_j \in \mathcal{V}^{S}\left(\mathcal{L}^\text{gNB}_{i}\right) \\
\textrm{PL}_{\mathrm{Out}}, &\text {otherwise},
\end{array}\right.
\end{equation}
where $\mathbb{P}_\text{LoS}$ and $\mathbb{P}_\text{NLoS}$ denote the probability of the link being in the LoS and NLoS state, respectively, and are dependent on the 2D distance $d_{i,j}^\text{2D}$~\cite[See Table 7.4.2-1]{3gpp2017study}. We emphasize that the $\mathbb{P}_\text{LoS}$ and $\mathbb{P}_\text{NLoS}$ are taken into account only when the UE is in the direct visibility region. For the points outside the direct visibility polygon of $\mathcal{L}^\text{gNB}_i$, $L_{ij}$ should be calculated using $\textrm{PL}_\text{NLoS}$, provided that the $\mathcal{L}_j^\text{SA}$ is in the indirect visibility region. We also note the NLoS paths can be blocked too but we assume that there exist at least a few MPCs that can be used to communicate with a SA in the indirect visibility region. 

Keeping the above perspective in mind, we characterize the path loss considering both the computational geometry and the statistical nature of the link between the gNB and SAs as:
\begin{align}\label{Eq:Final_PL}
    \textrm{PL}_{ij} &=  \; \delta_{ij} (\mathbb{P}_\text{LoS}(d^\text{2D}_{ij}) \textrm{PL}_\text{LoS}(d^\text{3D}_{ij}) + \mathbb{P}_\text{NLoS}(d^\text{2D}_{ij}) \textrm{PL}_\text{NLoS}(d^\text{3D}_{ij})) \nonumber \\ &+ (1 - \delta_{ij}) \times (\Delta_{ij} \textrm{PL}_\text{NLoS}(d^\text{3D}_{ij})) + (1-\Delta_{ij}) \textrm{PL}_\text{Out}),\hspace{-1cm}
    \end{align} 
which captures (\ref{Equ:PL_Dynamic}) using binary indicator functions $\delta_{ij}$ and $\Delta_{ij} \in \{0,1\}$, where $\delta_{ij} = 1,$ if the $\mathcal{L}^\text{SA}_j$ is inside the direct visibility region of $\mathcal{L}^\text{gNB}_i$, otherwise it is zero. Similarly, $\Delta_{ij} = 1$, if the $\mathcal{L}^\text{SA}_j$ is inside the indirect visibility region of $\mathcal{L}^\text{gNB}_i$ and it is zero otherwise. We note that the path loss models only capture the distance and frequency dependent factors and do not include the shadow fading~(SF) due to mathematical intractability owing to the stochastic nature. Including SF makes the optimization problem a stochastic problem, which can be solved using techniques such as chance-constrained optimization. However, these problems are often non-trivial and thus, we reserve this as a future work. For this work, we include the SF as a fading margin during the link budget calculation of $\gamma_{\max}$ threshold. We will elaborate on this discussion along with the path loss/ probability models in Section~\ref{Sim_Param_Sel}.\looseness=-1

\section{Deployment of mmWave {g}NBs}\label{Sec: gNB Placement}
Now that we have the path loss $\textrm{PL}_{ij}$ for each partition of the SA from the potential gNB locations, the goal now is to find the locations of a given number of gNB(s) $N^\text{gNB}$ that maximize the SA coverage. We address this goal by formulating the gNB maximum coverage optimization problem as a binary integer linear programming~(BILP) problem. 

\subsection{BILP Optimization Problem for mmWave {g}NBs}
We define $\alpha_i \in \{0,1\}$ to be a binary indicator variable which is 1 when the gNB at location $\mathcal{L}_i^\text{gNB}$ is selected to serve the SA, and it is zero otherwise. Similarly, let $\beta_j\in \{0,1\}$ be a coverage indicator which is 1 when the $\mathcal{L}_j^\text{SA}$ is covered, and it is zero otherwise. All the coverage related information is captured by the term $C_{ij} = \max\{0, \text{sign}(\gamma_\text{max} - \textrm{PL}_{ij})\}$. If $\textrm{PL}_{ij}~<~\gamma_\text{max}$, then the $C_{ij} = 1$, indicating the $\mathcal{L}_j^\text{SA}$ can be covered if the gNB is deployed at $\mathcal{L}_i^\text{gNB}$. If the QoS constraint $\gamma_\text{max} < \textrm{PL}_{ij}$ is not met, then $C_{ij}=0$.

The objective of the optimization problem is to maximize the weighted coverage of the SA under the arguments $\bm{\alpha}$ and $\bm{\beta}$ for a fixed number of gNBs $N^\text{gNB}$. With the above, the gNB weighted coverage optimization problem to find the optimal locations of $N^\text{gNB}$ gNBs over $M$ SAs can be formulated as~follows
\begin{eqnarray}\label{Eq:gNB_Coverage_Problem}
\begin{aligned}
 \underset{\bm{\alpha},\bm{\beta}}{\max} \quad & \sum_{j = 1 }^M w_j \beta_j\\
  \textrm{s.t.}  \quad & \textrm{c}_1: \quad \sum_{i=1}^N C_{ij} \alpha_i \geq \beta_j,\quad \forall j \in [M],\\
  &\textrm{c}_2: \quad \sum_{i=1}^N \alpha_i = N^\text{gNB},\\
  &\textrm{c}_3: \quad \alpha_i, \beta_j \in \{0,1\}, \quad  i \in [N], j \in [M]. 
 \end{aligned}
\end{eqnarray}
In~\eqref{Eq:gNB_Coverage_Problem}, the constraint c$_1$ states that the grid $\mathcal{L}_j^\text{SA}$ is covered only if the QoS constraint is met by at least one of the gNB(s). Constraint c$_2$ mandates the total number of gNBs to be equal to $N^\text{gNB}$, and the constraint c$_3$ states that both~$\bm{\alpha}$ and~$\bm{\beta}$ are binary indicator~variables. The term $w_j$ in the objective function denotes the normalized weight associated with each grid in the SA. The premise is that the UEs tend to be clustered around certain locations, such as bus stations, shops, etc., and those locations should be covered at any cost. On the other hand, if there is e.g. a large pond in the environment, then the area of the pond can be given a lower weight. One can think of other instances and contexts where the weights should be low or high. In general, the weights can be determined based on historical user-distribution data at the network provider or the features derived from the map. 

The choice of $\gamma_\text{max}$ in constraint c$_1$ depends on the link budget analysis, specifications of the gNB (vendor specifications), and the supported UEs capabilities (e.g., beamforming gains, gNB/UE cable loss, body loss, etc.). More details on the choice of $\gamma_\text{max}$ is discussed in Section~\ref{Section:Simulation}. The choice of the $\gamma_\text{max}$ also dictates the cell radius which gives us a heuristic to select the number of gNBs $N^\text{gNB}$. Let $a^\text{SA}$ and $a^\text{Cell} = \pi\times (\text{cell radius})^2$ correspond to the area of the SA and the area covered by a gNB, respectively. Then, the required number of gNBs can be estimated as $N^\text{gNB}~= ~\ceil{\frac{a^\text{SA}}{a^\text{Cell}}}$. The main idea here is to start with $N^\text{gNB}$ gNBs to place and then increase it as required. 

\subsection{BILP Solution for mmWave {g}NBs}
The BILP problem in~(\ref{Eq:gNB_Coverage_Problem}) is NP-Hard, and hence, solving it incurs a combinatorial complexity. The sub-optimal solutions for~(\ref{Eq:gNB_Coverage_Problem}) can be attempted via relaxation techniques (box constraints) or majorization-minimization techniques~\cite{MM_Approach_Integer,MMApproach_Yusuf}. However, relaxation methods may not provide optimal or even feasible solutions and may violate some of the constraints, but they will help to solve the problems in a polynomial time. 
Thus, we solve the BILP problem in~(\ref{Eq:gNB_Coverage_Problem}) by \textit{GUROBI-CVX} in MATLAB~\cite{cvx,gurobi}, one of the state-of-the-art BILP solvers, based on the branch-and-bound algorithm, which aims to search for the global optimum.

Solving (\ref{Eq:gNB_Coverage_Problem}) provides the optimum spatial locations (indices of $\alpha_i = 1$) for $N^\text{gNB}$ gNBs. {We denote the optimal locations of the $N^\text{gNB}$ gNBs as $\mathcal{L}^\text{gNB$_\star$}$}. Even though the determined set of gNB locations provide the maximum coverage area, some parts of the SA may still be in outage. 
These locations can be denoted by the set {$\mathcal{L}^{\text{OSA}}~\in ~\{(x_j^\text{SA},y_j^\text{SA},z_j^\text{SA})| \beta_j =~0, \forall j \in [M]\}$. The cardinality of $\mathcal{L}^{\text{OSA}}$ is denoted as $O$}, where $~O < ~M$. In the next section, our goal is to extend coverage to spatial locations in the set $\mathcal{L}^{\text{OSA}}$ using PMRs instead of increasing the number of gNBs $N^\text{gNB}$. Thus, from here onwards we restrict the $\mathcal{L}^\text{SA}$ to the set $\mathcal{L}^\text{OSA}$.

\section{Deployment of mmWave PMRs}\label{Sec:PMRs_Problem_Position_Orientation}
Deploying PMRs is an economical means to extend the coverage area, especially to NLoS areas. The critical questions are 1) how many reflectors need to be deployed, and 2) where should they be deployed, so that the coverage of the NLoS area is increased effectively. We address these questions in this section. We start by discussing the preliminaries of the PMRs and the physical constraints associated with their placement.




\begin{figure}[!t]
    \centerline{
    \includegraphics[trim=0cm 0cm 0cm 0.2cm, clip,scale=0.325]{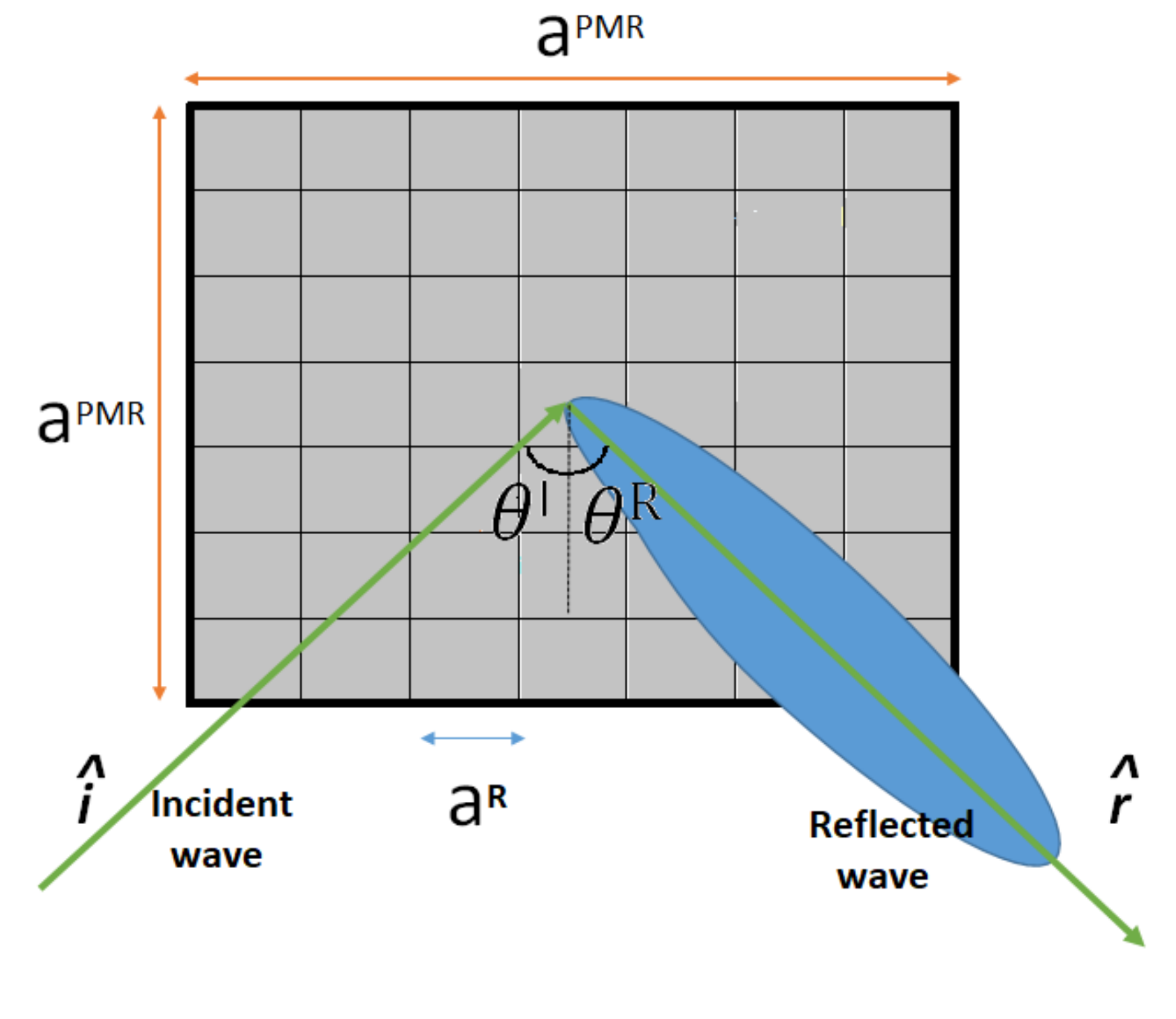}}
    \caption{A square reflector of side length $a^\text{PMR}$ with $R$ facets each of size $a^\text{R} \times a^\text{R}$.}
    \label{fig:Reflection}
\end{figure}

\begin{figure*}[!t]
    \begin{subfigure}{.6\textwidth}
    \includegraphics[scale=.35]{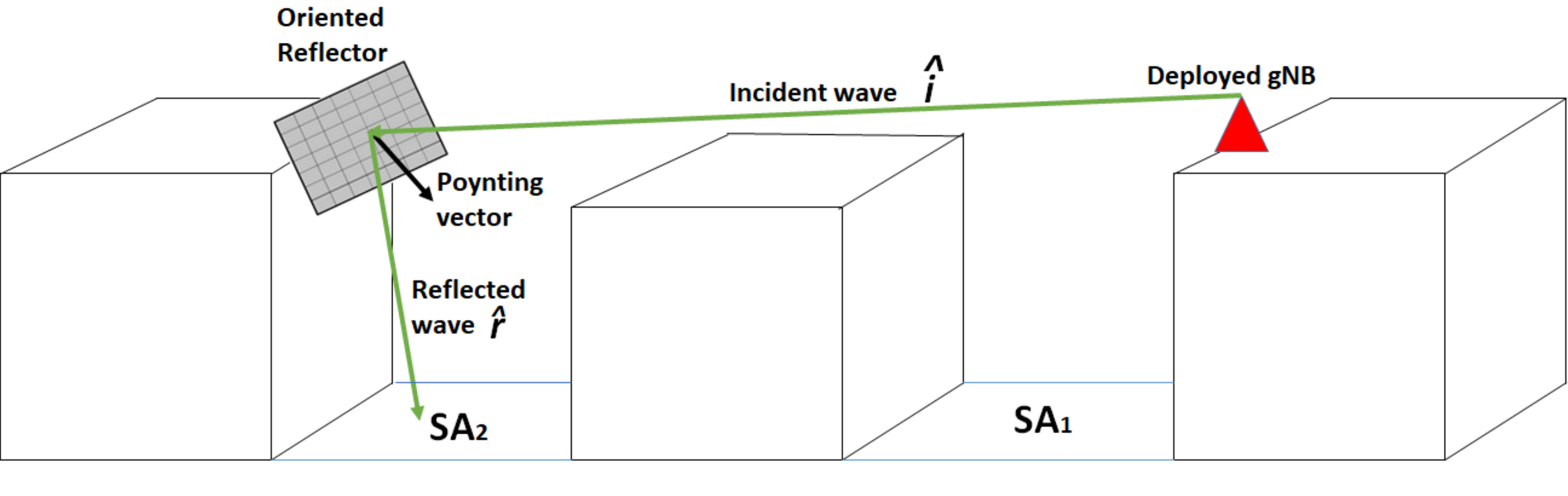}
    \caption{ }
    \end{subfigure}
    \begin{subfigure}{.24\textwidth}
    \centering
     \includegraphics[scale=0.275]{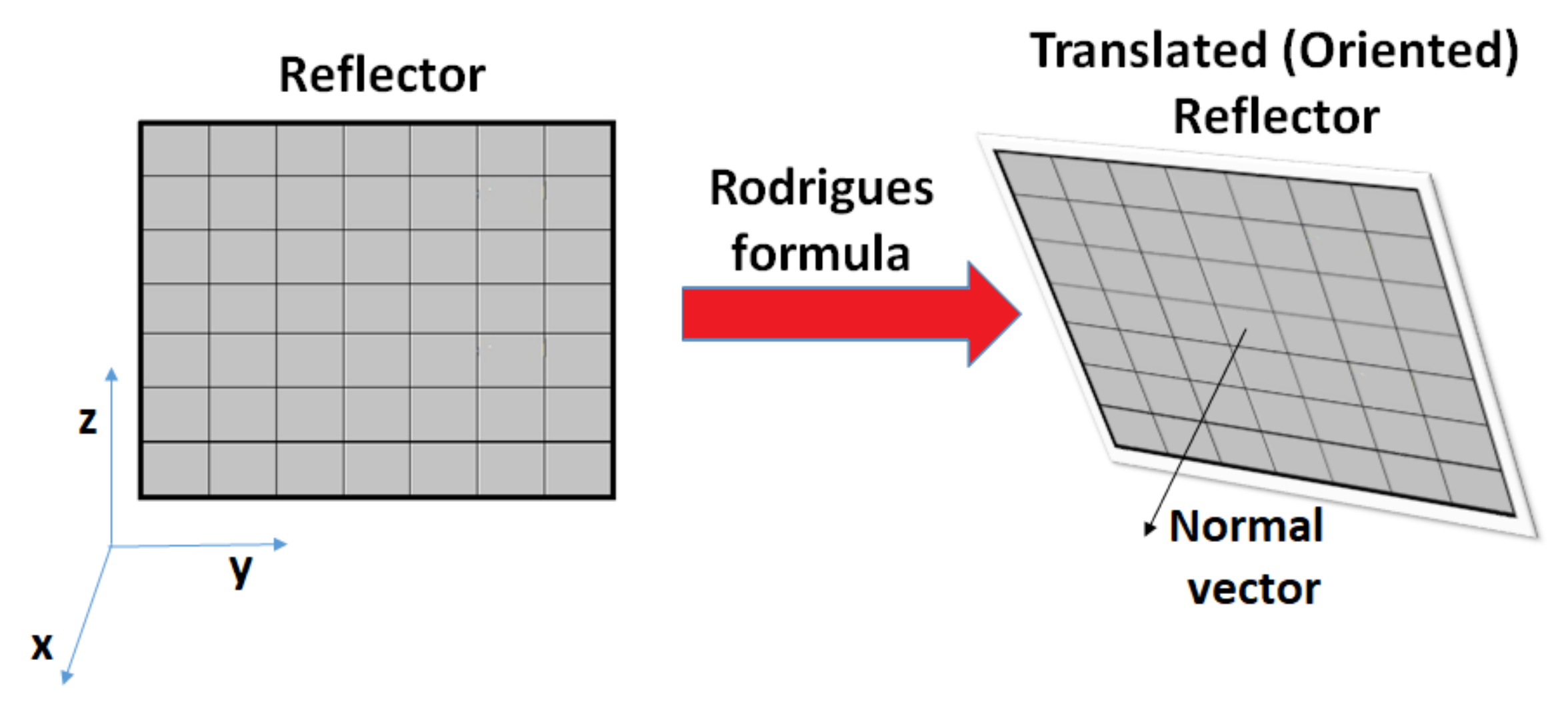}
     \caption{ }
    \end{subfigure}
    \caption{(a) Illustration of a simple scenario where the red triangle point is the position at which the gNB is deployed. The SA$_1$ is directly served via the gNB, whereas SA$_2$ can only be served by placing PMRs and orienting it properly to serve the SA. (b) Illustration of the PMR and translated points of the PMR.}
    \label{fig:Example_Scene}
    \vspace{-.5cm}
\end{figure*}
\subsection{Feasible PMR Deployment Locations}\label{Sec:Potential_PMR_Lcoations}
The feasible candidate locations for the PMRs can be defined as the locations that are visible from both the deployed gNB location(s) and the uncovered grids inside the NLoS area from the set $\mathcal{L}^\text{OSA}$. We denote the set of feasible candidate PMR locations by $\mathcal{L}^\text{PMR}$ and its cardinality as $K$, and those locations can be mathematically represented~as
\begin{eqnarray}\label{Eq:PMR_Candidates_Buildings}
\begin{aligned}
& \mathcal{L}^\text{PMR} =  \mathcal{L}^{{\text{VgNB$_\star$}}}  \bigcap \mathcal{L}^{{\text{VOSA}}}  \bigcap \mathcal{L}^\text{Build}, 
\end{aligned}
\end{eqnarray}
where,
\begin{eqnarray}
\begin{aligned}
\mathcal{L}^{{\text{VgNB$_\star$}}}  &= \Big\{ {\textstyle \bigcup\limits_i}\mathcal{V}(\mathcal{L}^\text{gNB$_\star$}_i ) \mathlarger{\mid} \;   \forall i \in [N^\text{gNB}]\Big\},\\
\mathcal{L}^{{\text{VOSA}}} & = \Big\{ {\textstyle \bigcup\limits_j} \mathcal{V}(\mathcal{L}^\text{OSA}_j)\mathlarger{\mid} \; \forall j \in [O] \Big\}
.
\end{aligned}
\end{eqnarray}
The set $\mathcal{L}^{{\text{VgNB$_\star$}}}$ contain the set of all the points visible from gNB locations obtained from (\ref{Eq:gNB_Coverage_Problem}). Similarly, $\mathcal{L}^{{\text{VOSA}}}$ contain the set of points visible from the outage SA $\mathcal{L}^\text{OSA}$. 
The set $\mathcal{L}^\text{Build}$ contains the 3D coordinates of $B$ points on the outer surfaces of the buildings (see~\eqref{eq:C_build}) as discussed in Section~\ref{Sec:PMR_Loc}. Note that (\ref{eq:C_build}) denotes the potential PMR locations without considering the NLoS areas and gNB locations. Whereas, (\ref{Eq:PMR_Candidates_Buildings}) reduces the potential locations (by construction $K < B$) by considering the NLoS areas and determined gNB locations. 
Up next, we present the preliminaries and physical constraint associated with placing a PMR.

\subsection{PMRs: Preliminaries and Physical Constraints}
Let the unit directional vector of the incidence ray (from the gNB to the reflector) and the reflected ray (from the reflector to the desired spatial location) are denoted by $\hat{\textbf{i}}$ and $\hat{\textbf{r}}$, respectively. Also, let $\hat{\textbf{n}}$ denote the Poynting normal vector pointing outwards to the PMR, forming an angle of $\theta^\text{I}$ and~$\theta^\text{R}$ between the incident and reflected ray, respectively, as illustrated in Fig.~\ref{fig:Reflection}. According to Snell's law, the reflection takes place with an equal angle of incidence and reflection, that is, $\theta^\text{I} = \theta^\text{R}$, around the point of impact with respect to the normal vector. \looseness = -1

Based on the above, the reflected ray can be represented in terms of the incident ray and the normal vector as~\cite{de2006reflections}:
\begin{equation}\label{Eq:Reflected_Ray}
\hat{\mathbf{r}} = \hat{\mathbf{i}} - 2 \left( \hat{\mathbf{i}} \cdot \hat{\mathbf{n}}\right) \hat{\mathbf{n}} = \hat{\mathbf{i}} - 2 \cos(\theta^\text{I}).
\end{equation}
which implies that the direction of a reflected ray (maximum response), and hence the receive angle at a particular location $\mathcal{L}_j^\text{OSA}$,
depends on the direction of the incident wave and the normal vector (orientation) of the reflector. Thus, we have two degrees of freedom for deploying the PMRs: the position at which the reflector is placed and the direction at which it is oriented. With the gNB and reflector positions fixed, the orientation of the reflector controls the angle of incidence and reflection, which, in turn, controls the direction of the reflected ray. This phenomenon is illustrated with an example scenario up next. \looseness =-1

\subsection{PMR Position and Orientation}
Let us consider the scenario in Fig.~\ref{fig:Example_Scene}(a) where SA$_1$ is in direct visibility to gNB. On the other hand, the SA$_2$ is not directly visible to the gNB. By properly deploying a PMR, the SA$_2$ can be made to be indirectly visibile to the gNB so that the communication can happen through first order reflections. Assume that the desired location to be covered is $\mathcal{L}_j^\text{OSA}$ and we have the gNB location(s) fixed using (\ref{Eq:gNB_Coverage_Problem}). Then the reflector position and orientation can be determined as follows.

\subsubsection{Reflector Position}
For analysis, we divide the whole PMR into a total of $R$ smaller size facets, each of size $a^\text{R} \times a^\text{R}$, and use the centroid of the reflector as our reference (used interchangeably as the position of the reflector) as illustrated in Fig~\ref{fig:Reflection}. Each of these facets is indexed by $r \in~[R]$. For simplicity, we fix the reflector position at a location $\mathcal{L}^\text{PMR}_k$ and calculate the reflector orientation as shown next.

\subsubsection{Reflector Orientation} With the gNB location and reflector location fixed, the unit directional incident ray $\hat{\mathbf{i}}_{i,k}$ from the $\mathcal{L}_i^\text{gNB$_\star$}$ to the reflector positioned at $\mathcal{L}_k^\text{PMR}$ for the centroid facet can be obtained by considering their Cartesian coordinates as $\hat{\mathbf{i}}_{i,k} = (\mathcal{L}^\text{gNB$_\star$}_{i} - \mathcal{L}^\text{PMR}_k)/ d^\text{3D}_{i,k}$. 
We ignore the index of the facet to avoid confusion while calculating the~orientation. Likewise, if the target area that needs to receive the maximum response is fixed, say $\mathcal{L}_j^\text{OSA}$, then the unit directional reflected ray from the reflector positioned at $\mathcal{L}_k^\text{PMR}$ can be calculated as $\hat{\mathbf{r}}_{j,k} = (\mathcal{L}^\text{OSA}_j- \mathcal{L}^\text{PMR}_k)/ d^\text{3D}_{j,k}$. Thus, with the known unit directional incident and reflected ray, the normal vector (orientation or Poynting vector) can be obtained as \begin{equation}\label{Eq:Snells_law}
    \hat{\mathbf{n}}_{i,j,k} = \frac{\hat{\mathbf{i}}_{i,k} - \hat{\mathbf{r}}_{j,k} }{2\left( \cos(\theta_{i,k}^\text{I})\right)}.
\end{equation}

The orientation in \eqref{Eq:Snells_law} at the reflector positioned at $\mathcal{L}^\text{PMR}_k$ ensures that there exists a ray that can reach the desired location $\mathcal{L}_j^\text{OSA}$, provided that an MPC hits the centroid facet of the reflector from the gNB $\mathcal{L}_i^\text{gNB$_\star$}$. Understanding this simple scheme is also the first step toward understanding the coverage area associated with all the facets. There are also other facets of the reflector which may cover other spatial locations. To determine this, we fix the orientation of the reflector as obtained in (\ref{Eq:Snells_law}) by performing the rotation of the solid reflector at $\mathcal{L}^\text{PMR}_k$ via the standard \textit{Euler-Rodrigues rotation formula}~\cite{dai2015euler} for all the facets (the whole reflector); for brevity, we skip the technical details which can found in any standard computer vision textbook (e.g.~\cite{szeliski2010computer,murray1994mathematical}). The translated reflector is illustrated in Fig.~\ref{fig:Example_Scene}(b). 



\subsection{Reflector Path Loss Calculation}
For calculating the path loss through a PMR, we adopt the model from~\cite{ozdogan2019intelligent}, which is derived using the standard physics-based electromagnetic plate scattering theory. 
The path loss (in linear scale) between the $\mathcal{L}_i^\text{gNB$_\star$}$ and $\mathcal{L}_j^\text{OSA}$ through the PMR positioned at $\mathcal{L}_k^\text{PMR}$ and {oriented for $\mathcal{L}_l^\text{OSA}$ (receives maximum response)} via the $r^\text{th}$ facet can be calculated as
\begin{equation}
\label{PMR_PathLoss_Model}
\textrm{PL}^{\text{PMR}}_{i,j,k,l,r} =  G_\text{gNB} G_\text{UE} \frac{(a^\text{R})^4 \cos^2(\theta^\text{I}_{i,k,l,r})}{(4\pi)^2(d^\text{3D}_{i,k,l,r} d^\text{3D}_{i,j,k,l,r})^\zeta } 
    \eta_{i,j,k,l,r},
\end{equation}
where
\begin{equation}\label{Eq:Sinc_Term}
    \eta_{i,j,k,l,r}= \left(\frac{\sin(\frac{\pi a^\text{R}}{\lambda} (\sin(\theta^R_{i,j,k,l,r}) - \sin(\theta^\text{I}_{i,k,l,r})))}{\frac{\pi a^\text{R}}{\lambda} (\sin(\theta^R_{i,j,k,l,r}) - \sin(\theta^\text{I}_{i,k,l,r}))} \right)^2,
\end{equation}
where $d^\text{3D}_{i,k,l,r}$ and $\theta^\text{I}_{i,k,l,r}$ are the distance and the incident angle from the $\mathcal{L}_i^\text{gNB$_\star$}$ to the $r^\text{th}$ facet of the reflector positioned at $\mathcal{L}^\text{PMR}_k$, oriented for $\mathcal{L}_l^\text{SA}$. Similarly, $d^\text{3D}_{i,j,k,l,r}$ and $\theta^R_{i,j,k,l,r}$ are the distance and the reflected angle from the $r^\text{th}$ facet of the reflector (positioned at $\mathcal{L}_k^\text{PMR}$, oriented for $\mathcal{L}_l^\text{OSA}$, associated with the $\mathcal{L}_i^\text{gNB$_\star$}$) to the $\mathcal{L}_j^\text{OSA}$. The terms $G_\text{gNB}$, $G_\text{UE}$, $\zeta$, and $\lambda$ denote the directional gain at the gNB and UE, path loss exponent, and wavelength of operation respectively.

Unlike the standard path loss models, the path loss through a PMR in~\eqref{PMR_PathLoss_Model}
depends on the area of the reflector, as well as the incidence and the  reflected wave angles. Earlier work, such as~\cite{peng2015effective}, have assumed specular reflection from the metallic reflectors at the wireless bands. However, this is dis-conjectured in~\cite{ozdogan2019intelligent}, where a finite-sized PMR can provide almost specular reflection in the visible spectrum, but 4-5 orders-of-magnitude wider beamwidths in the typical radio bands. This phenomenon is captured by the squared sinc term in (\ref{Eq:Sinc_Term}) and illustrated in Fig.~\ref{fig:Reflection} by the blue diffuse-scattered beam. 
\begin{figure*}[!t]
\begin{subfigure}{.33\textwidth}
\centering
\includegraphics[scale=0.4]{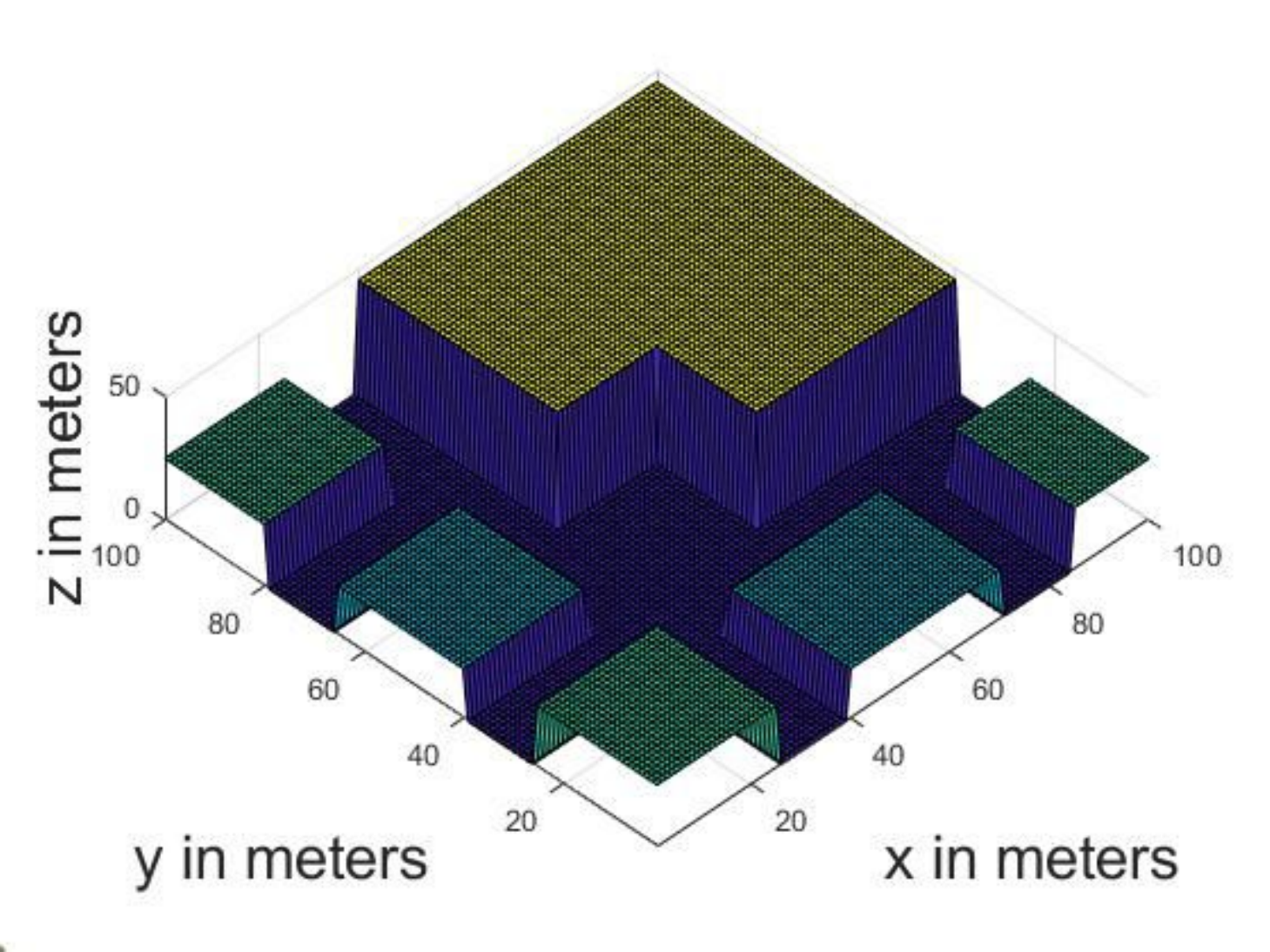}
\caption{ }
\end{subfigure}
\begin{subfigure}{.33\textwidth}
\centering
\includegraphics[scale=0.3]{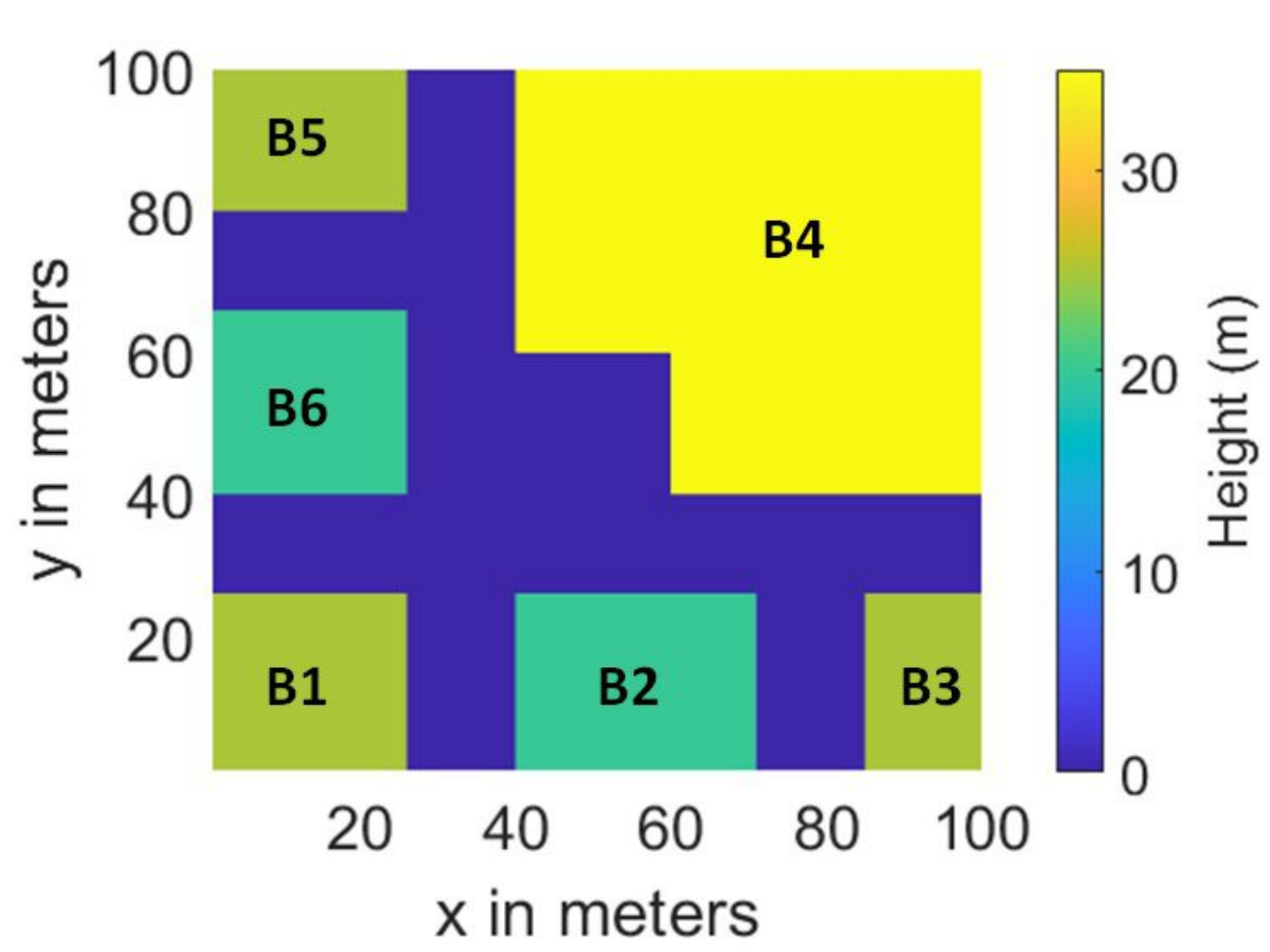}
\caption{ }
\end{subfigure}
\begin{subfigure}{.33\textwidth}
\centering
\includegraphics[scale=0.38]{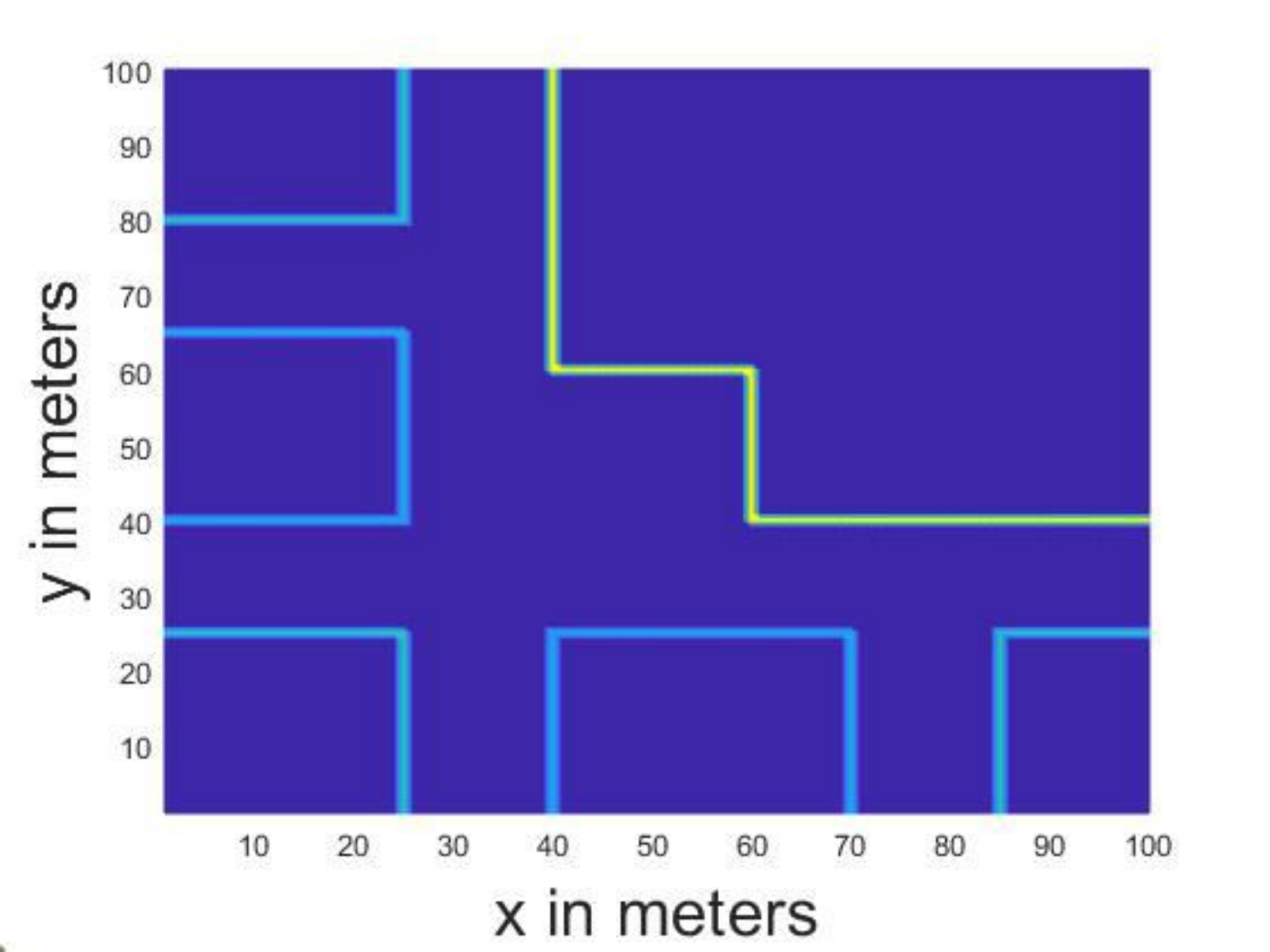}
\caption{ }
\end{subfigure}
\caption{(a) DEM of the considered 3D scenario. The blue points correspond to the SA $\mathcal{L}^\text{SA}$ and the remaining colored points correspond to the building's rooftop, respectively. (b) 2D view of the DEM (c). Boundary points of the rooftop of the buildings corresponds to the candidate locations of the gNBs~$\mathcal{L}^\text{gNB}$.}
\label{Fig:Scenario}
\end{figure*}
Finally, the total path loss (total power received/ power at the gNB) at the $\mathcal{L}_j^\text{OSA}$ from the reflector positioned at $\mathcal{L}_k^\text{PMR}$, oriented for $\mathcal{L}_l^\text{OSA}$, and associated with $\mathcal{L}_i^\text{gNB}$, is the sum of path loss from all the $R$ facets of the reflector under the idealized constructive interference. This is mathematically represented as $\textrm{PL}^{\text{PMR}}_{i,j,k,l}= \sum_{r=1}^R \textrm{PL}^{\text{PMR}}_{i,j,k,l,r}$. A theoretical justification of the above procedure is that, for the PMR, it does not matter if the total area is made up of many small or a few large plates, the maximum received power is the same and captured by the path loss model~\cite{ozdogan2019intelligent,khawaja2020multiple}. We exhaustively search for different reflector positions $\mathcal{L}^\text{PMR}_k$, and orientations to each $\mathcal{L}_l^\text{OSA}$ to populate $L^{\text{PMR}}_{i,j,k,l}$.

\subsection{PMR BILP Optimization Problem}
Similar to the gNB placement problem, we formulate and solve the PMR placement problem as a BILP optimization problem as follows: 
\begin{eqnarray}\label{Eq:PMR_Optimization}
\begin{aligned} 
  \underset{\bm{\alpha}, \bm{\beta}, \bm{\xi}}{\max} & \quad \sum_{j = 1 }^O  w_j \beta_j\\
   \textrm{s.t.}  \\
 \quad \textrm{c}_1: & \quad \sum_{i=1}^{N^\text{gNB}} \sum_{k=1}^K \sum_{l = 1}^{O} \textrm{PL}^\text{PMR}_{i,j,k,l} \alpha_{i,k,l} = \xi_j,\quad \forall j \in [O],\\
   \quad \textrm{c}_2: & \quad M_b(\beta_j - 1)  \leq \xi_j - \gamma_{\max} \leq M_b \beta_j,\quad \forall j \in [O],\\
   \quad \textrm{c}_3: & \quad   \sum_{i=1}^{N^\text{gNB}} \sum_{k=1}^K \sum_{l=1}^O  \alpha_{i,k,l} = N^\text{PMR},\\
 \quad \textrm{c}_4: & \quad \sum_{k=1}^K\sum_{j=1}^O \alpha_{i,k,l}  \leq 1, \quad \forall i \in [N^\text{gNB}], \\
 \quad \textrm{c}_5: & \quad \sum_{l=1}^O \alpha_{i,k,l} \leq 1, \quad \forall i \in [N^\text{gNB}], k \in [K],\\
 \quad \textrm{c}_6: & \quad \beta_j,  \in \{0,1\}, \quad \forall j \in[O],\\
 \quad \textrm{c}_7: & \quad \alpha_{i,k,l} \in \{0,1\}, \quad \forall i \in [N^\text{gNB}], k \in [K],  l \in [O].
\end{aligned}
\end{eqnarray}
In the above optimization problem, the goal is to find
the best locations/orientations of a given number of PMRs $N^\text{PMR}$ that maximize the SA coverage. The term $\alpha_{i,k,l} \in~\{0,1\}$ is a binary indicator variable which is~1 when the reflector positioned at $\mathcal{L}^\text{PMR}_k$, oriented for $\mathcal{L}_l^\text{OSA}$, and associated with $\mathcal{L}_i^\text{gNB$_\star$}$ is serving the NLoS area, otherwise it is zero. Likewise, $\beta_j\in \{0,1\}$ is a coverage indicator which is 1 when the $\mathcal{L}_j^\text{OSA}$ is covered, otherwise it is zero. The term $w_j$ denotes the normalized weight associated with each grid in the $\mathcal{L}^\text{OSA}$. With this defined, the goal of the optimization problem is to maximize the weighted coverage (for the set $\mathcal{L}^\text{OSA}$) for a fixed number of reflectors denoted by $N^\text{PMR}$.


In~\eqref{Eq:PMR_Optimization}, the linear constraint c$_1$ captures the total power at $\mathcal{L}^\text{OSA}_j$ for the choice of the reflector placement $\alpha_{i,k,l}$ which meets the QoS constraint. The QoS constraint ($\gamma_{\max}$ is in linear scale) is captured with the help of the Big-$M$ formulation~\cite{newman2013survey} in constraint c$_2$, where $M_b$ is a small-but-sufficiently-large constant. If $M_b$ is smaller than the possible solution of $\xi_j - \gamma_{\max}$, the optimization problem may cut off valid (maybe optimal) solutions. If $M_b$ is too large, the model may become numerically difficult due to ill-conditioned problem. For simulations, we set $M_b = \sum_{i,k,l}\textrm{PL}_{i,j,k,l}^\text{PMR}$, which is the maximum value that can be attained (the upper bound). As an example, whenever the total power $\xi_j$ at $\mathcal{L}^\text{OSA}_j$ exceeds the QoS power constraint $\xi_j - \gamma_{\max} > 0$, the big-$M_b$ formulation forces the indicator $\beta_j = 1$ satisfying the QoS constraint. The $\beta_j$ term will be zero ($\xi_j - \gamma_{\max} < 0$) when the QoS constraint is not met. \looseness =-1 

The constraint c$_3$ mandates that the number of reflectors to be placed is restricted to $N^\text{PMR}$ reflectors. The number of reflectors is chosen to be a free parameter and is varied in the simulation. The constraints c$_4$ and c$_5$ force the optimization problem to choose the reflectors such that only one possible orientation should be selected per position, and each reflector should be associated with at most one gNB. Finally, the indices of the term $\alpha_{i,k,l} = 1$ correspond to the spatial locations/orientations where the reflectors are to be placed to serve the NLoS area. Similar to the gNB deployment problem, the BILP problem in~\eqref{Eq:PMR_Optimization} is solved with the help of the \textit{CVX-Gurobi} solver~\cite{cvx,gurobi}.

\section{Numerical Results}\label{Section:Simulation}
In this section, we demonstrate the efficacy of the proposed approach with the help of an example 3D map. We consider a UMa outdoor-to-outdoor communication scenario, where the gNBs can be placed on the rooftop of the buildings at a height ($z^\text{gNB}$) of 25 m to 30 m from the ground. The UEs are the outdoor ground users at a height ($z^\text{SA}$) of 1.5 m. The DEM of the example map is shown in Fig.~\ref{Fig:Scenario}(a). This map is considered for the ease of reconstruction of the same scenario in Remcom Wireless InSite ray tracing simulator~\cite{WirelessInsite}, whose output can serve as a comparison\footnote{{Matlab script to generate the results and the  ray tracing data are available online at: \url{https://research.ece.ncsu.edu/
mpact/data-management/}}}. We stress that the choice of a simple map is biased in favor of showing the proof of concept. 

The considered scenario consists of six buildings of different heights, namely B1 through B6 with heights of 25~m, 20~m, 25~m, 35~m, 25~m, and 20~m, respectively, starting from the (0,0,0) and moving in the counter-clockwise direction. The 2D view of the same along with the building widths are shown in Fig.~\ref{Fig:Scenario}(b). As noted in Section~\ref{Potential_gNB_Locations}, the rooftop boundaries of the buildings are the candidate gNB locations and shown in Fig.~\ref{Fig:Scenario}(c). We maintain a resolution of 1~m $\times$ 1~m for the gNB candidates resulting in a total of $N = 418$ potential gNB locations $\mathcal{L}^\text{gNB}$. As noted before, one can reduce the number of potential candidate points by means of several techniques, such as viewshed analysis~\cite{fatih2020MABP}. In this work, we consider all the potential gNB points. The blue points in Fig. \ref{Fig:Scenario}(b) correspond to the partitioned SAs $\mathcal{L}^\text{SA}$, a total of $M = 3704$ grids, where each grid is of size 1~m $\times$ 1~m.

\begin{table}[!t]
\centering
\renewcommand{\arraystretch}{1.6}
\caption{Channel and Link State Probability Models.}
\begin{tabular}{|P{2cm}|P{5.85cm}|}
\hline
\textbf{Parameter}& \textbf{Model}    \\
\hline
$\textrm{PL}_\text{LoS}(d^\text{3D},f^\text{C})$ & $28 + 22\log_{10}(d^\text{3D}) + 20 \log_{10}(f^\text{C})$ \\ \hline
$\textrm{PL}_\text{NLoS}(d^\text{3D},f^\text{C})$ & $  \max (\textrm{PL}_\text{LoS}(d^\text{3D}, 13.54 + 39.08\log_{10}(d^\text{3D}) + 20\log_{10}(f^\text{C}) - 0.6 (z^\text{SA} - 1.5) )$ \\ \hline
$\mathbb{P}_\text{LoS}$ & \text{\footnotesize $\min(\frac{18}{d^\text{2D}},1) \left(1+ \exp(-\frac{d^\text{2D}}{63})\right) + \exp\left(\frac{-d^\text{2D}}{63}\right)$} \\ \hline
\end{tabular}
\label{Table_Sim}
\end{table}

\subsection{Choice of Simulation Parameters}\label{Sim_Param_Sel}
For the path loss model, we adopt the 3GPP TR 38.901~\cite{3gpp2017study}[Table 7.4.1-1: Path loss models] omni-directional UMa LoS and NLoS path loss models, which are shown in Table~\ref{Table_Sim}. The $\textrm{PL}_\text{LoS}$ model is chosen such that $d^\text{2D}$ is within the break-point distance ($d^\text{BP}$) threshold regime, i.e., $d^\text{BP} \geq d^\text{2D}$ (which holds true for the considered scenario), where \text{$d^\text{BP} = 4 (z^\text{gNB} -1)(z^\text{SA} -1) f^\text{C} 10^9 / c$}, $c$ is the speed of light, and $f^\text{C}$ is the carrier frequency of operation (in GHz). See~\cite{3gpp2017study}[Table 7.4.1-1: Path loss models] for more details. 

\begin{figure*}[!t]
\begin{subfigure}{.33\textwidth}
\centerline{\includegraphics[scale=0.27]{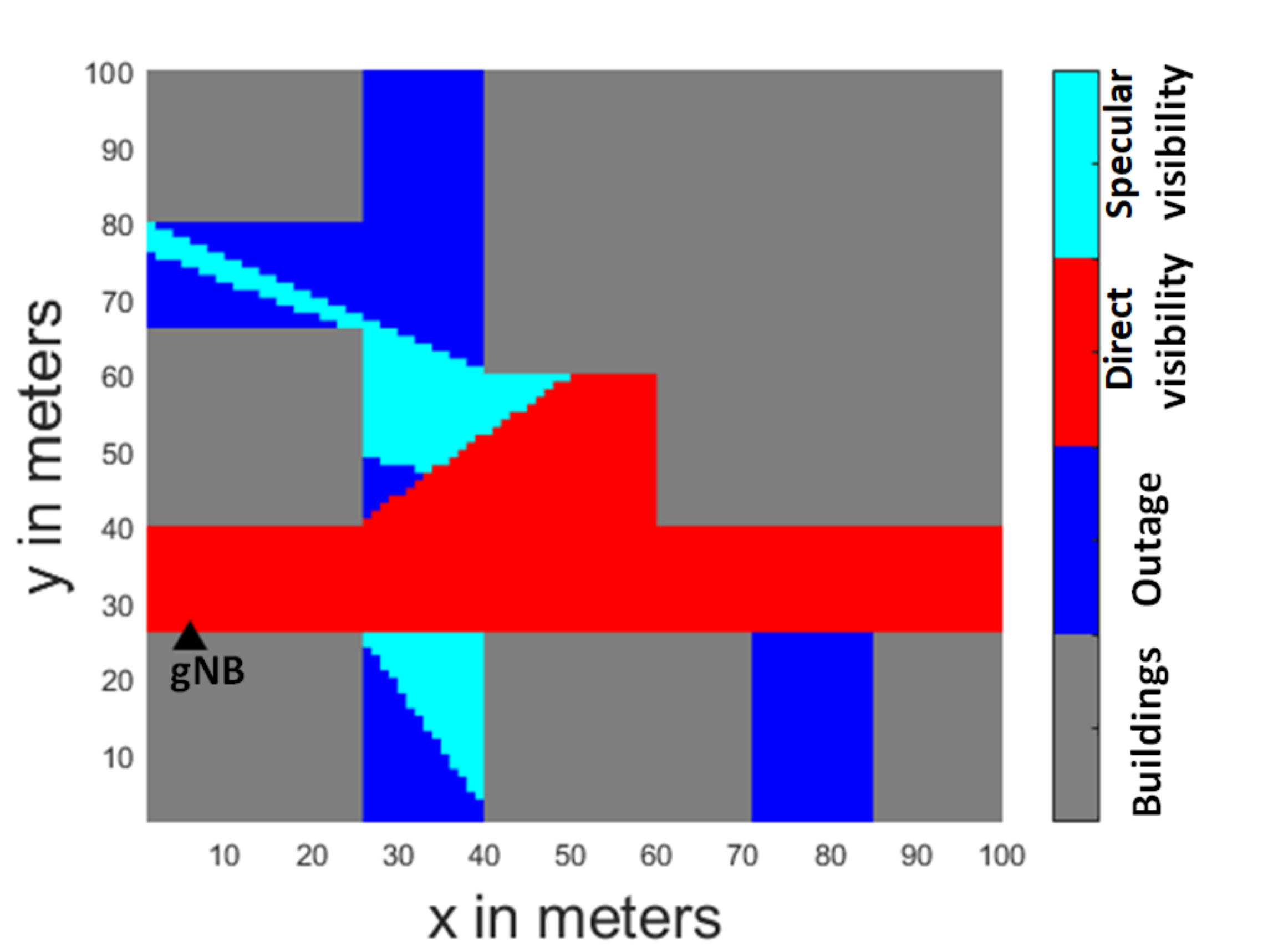}}
\caption{ }
\end{subfigure}
\begin{subfigure}{.33\textwidth}
\centerline{\includegraphics[scale=0.27]{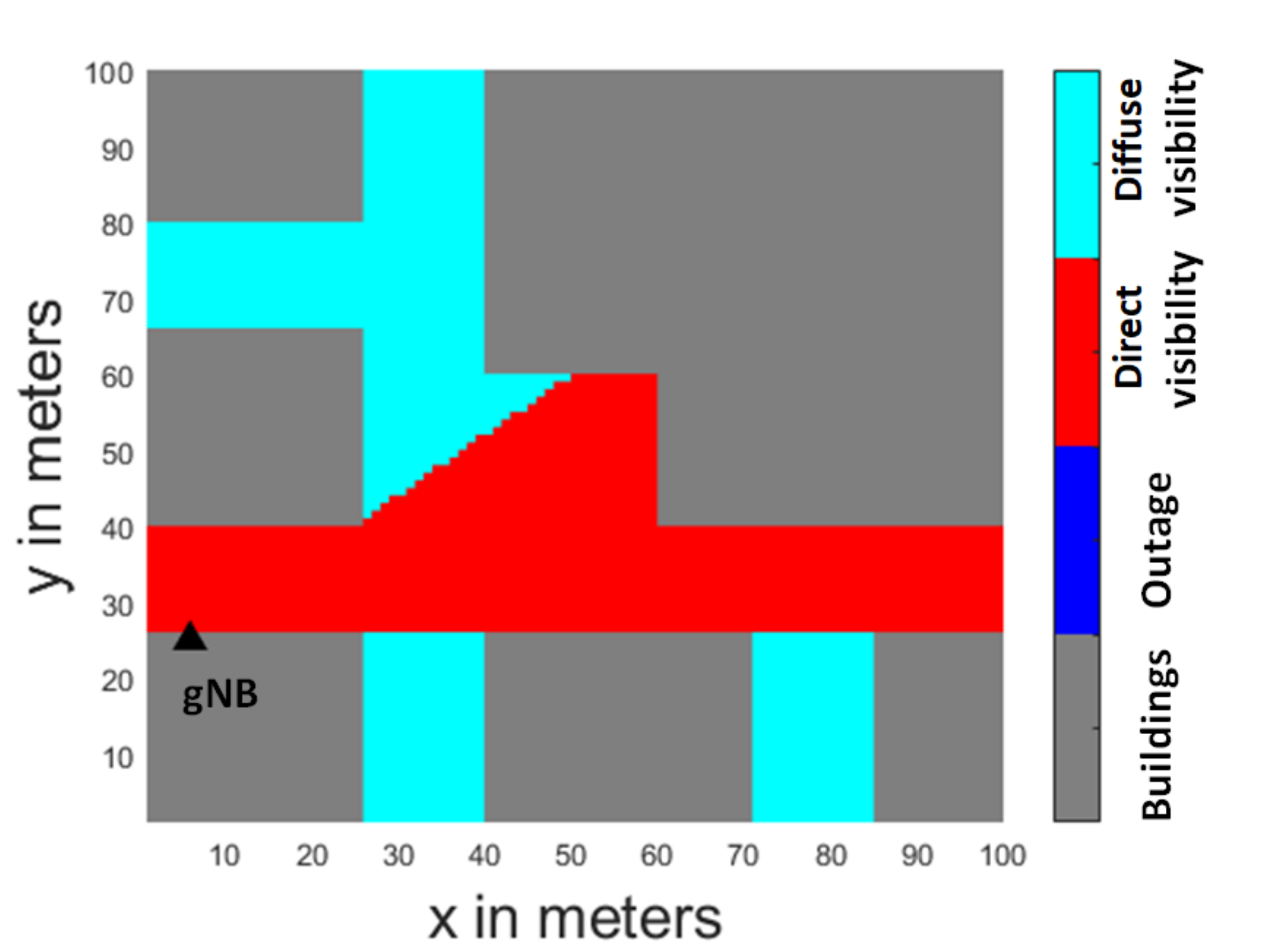}}
\caption{ }
\end{subfigure}
\begin{subfigure}{.33\textwidth}
    \centerline{\includegraphics[scale=0.365]{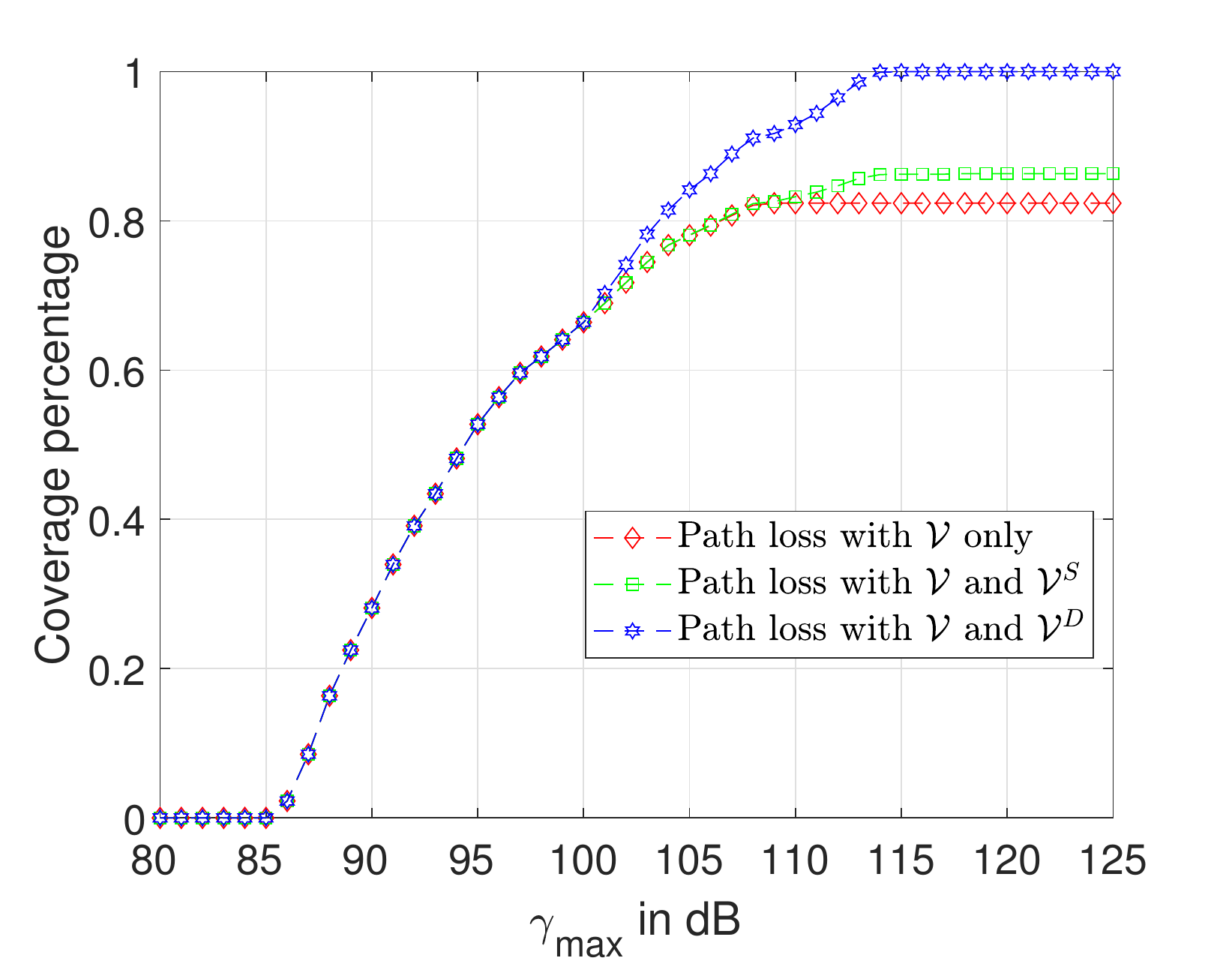}}
    \label{fig:Coverage_Vs_MAPL}
\caption{ }
\end{subfigure}
\caption{Illustration of the (a) specular and (b) diffuse visibility regions for a fixed gNB location: The black triangle denotes the gNB location selected randomly among the set of potential gNB locations. 
(c) The fraction of SA covered by the gNB as a function of the MAPL threshold $\gamma_{\max}$ under different scenarios considering the direct visibility, direct and specular visibility, and direct and diffuse visibility alone.}
\label{Fig:Visibility_Illustration}
\vspace{-.5cm}
\end{figure*}

The LoS probability for the UMa scenario is chosen as shown in Table~\ref{Table_Sim}. We operate in the regime where $z^\text{SA}$ is less than~13 m (\cite[See Table 7.4.2-1]{3gpp2017study}). Finally, the NLoS probability $\mathbb{P}_\text{NLoS}$ is given by 1 - $\mathbb{P}_\text{LoS}$. We note that one can utilize other notable blockage models such as 3GPP $d_1/d_2$, NYU squared, and inverse exponential models discussed in~\cite{sun2015path}. However, in this work, we restrict our discussion and simulation to the 3GPP UMa scenario model.

\begin{table}[!t]
\renewcommand{\arraystretch}{1.15}
\centering
\caption{Typical values of gains/losses and summary.}
\label{Table:Table1}
\begin{tabular}{|P{1.5cm}|P{1.5cm}|P{4.5cm}|}
\hline
\textbf{Parameter}& \textbf{Value} & \textbf{Meaning}   \\
\hline
$f^C$ &  28 GHz & mmWave frequency band \\ \hline
$W$ & 100 MHz & Downlink bandwidth \\ \hline
$P_\text{gNB}$ & 49 dBm & Maximum transmit power at the gNB\\  \hline
$G_\text{gNB}$ & 21.5 dBi & Antenna gain at the gNB \\ \hline
$G_\text{UE}$ & 5.5 dBi & Antenna gain at the UE \\  \hline
$L_\text{cab}$ &  2 dB & gNB cable loss \\ \hline
$L_\text{body}$ &  13 dB & Body loss \\ \hline
$L_\text{fol}$ &  16 dB & Foliage loss \\ \hline
$L_\text{rain/ice}$ &  3 dB & Rain/Ice loss margin \\ \hline
$L_\text{int}$ &  1 dB & Interference margin \\ \hline
$L_\text{SF}$ &  7 dB & Slow fading margin \\ \hline
$L_\text{oth}$ & 3 dB & Additional losses (such as sub-carrier quantity, penetration losses, etc) \\ \hline
$N_0$  &  -94 dBm & Thermal noise (-174 + 10$\log_{10}(W)$) \\ \hline
$NF_\text{UE}$ &  5 dB & UE noise figure \\ \hline
\end{tabular}
\label{Table1}
\end{table}

The choice of maximum allowable path loss~(MAPL)~$\gamma_{\max}$ can be obtained from the 5G NR budget analysis~\cite{5G_Budget1,5G_Budget2}. The MAPL~$\gamma_{\max}$ (in dB) can be linked to the total gains/losses in the system and the desired demodulation threshold signal-to-interference ratio (SINR) as follows
\begin{eqnarray}
\begin{aligned}
     & \gamma_\text{max} = {P}_\text{gNB} + {G}_\text{tot} - {L}_\text{tot} - {R}_\text{sens},
    \end{aligned}
\end{eqnarray}
where the total gain/losses in the system is composed of ${G}_\text{tot} = G_\text{gNB} + G_\text{UE}$ and ${L}_\text{tot} = {L}_\text{cab.} - {L}_\text{body} - {L}_\text{fol.} - {L}_\text{rain/ice}  - {L}_\text{int.} - {L}_\text{SF} - L_\text{oth.}$, respectively. Further, the receive sensitivity at the UE is ${R}_\text{sens} = N_0 + {NF}_\text{UE} + {SINR}$. The meaning and the typical values of the above parameters are provided in Table~\ref{Table1}. We emphasize that these parameters are vendor specific and vary with specifications. The major part of the MAPL is the desired SINR, meaning higher the desired SINR values (better modulation) lower the MAPL, implying smaller cell range. With the tabulated parameters, we have $\gamma_\text{max} = 121 - SINR$ dB. In the simulations, we vary $\gamma_{\max}$ from $85$~dB to $125$~dB. As the cell range varies, we fix the number of gNBs $N^\text{gNB}$ to be equal to one for a fair comparison of coverage results for the considered pilot problem. We note that our proposed method is amiable with multiple BSs, as the network densification is one of the key characteristics of 5G communications in mmWave band. Next, we look into the results of GB-PLM and gNB deployment problem followed by the PMR results in Section~\ref{Sec:PMR_Results}.

\subsection{GB-PLM and gNB Deployment}
For solving (\ref{Eq:gNB_Coverage_Problem}), we assume all the grids in the SA are equally favorable (i.e., have equal weights) and choose other simulation-free parameters as shown in Table~\ref{Table_Sim}.

\subsubsection{Direct and Indirect Visibility Calculation} Section~\ref{Sec:Visibility_PL_Model} dealt with definitions rather than constructions of direct and indirect visibility regions. Fig.~\ref{Fig:Visibility_Illustration}(a)-(b) illustrates such constructions for the considered 3D map. Fig.~\ref{Fig:Visibility_Illustration}(a) shows the direct (red-colored grids) and specular (cyan-colored grids) reflection regions for a fixed gNB location. The blue points are the grids that cannot be reached by the gNB through static scatterers. 
Similarly, Fig.~\ref{Fig:Visibility_Illustration}(b) illustrates the direct and diffuse (cyan-colored SAs) reflection regions from the same fixed gNB location. \looseness =-1 

For both the specular and diffuse region construction, we enabled all the mirror edges (building scatterers) assuming all the scatterers are glass-based structures or concrete-based structures for the specular and the diffuse regions, respectively. As seen from both figures, it is evident that the specular visibility region is always a subset of the diffuse visible region. In fact, the diffuse visibility region can be viewed as a region that captures the visibility from the dynamic elements due to its inherent construction. This direct and indirect visibility calculation helps us to populate $\delta_{ij}$ and $\Delta_{ij}$ in~\eqref{Eq:Final_PL} for all the considered points.

\subsubsection{Fraction of Coverage from gNB}
With the above specified parameter settings, solving (\ref{Eq:gNB_Coverage_Problem}) gives the optimum gNB locations for maximum coverage in the SA. Fig.~\ref{Fig:Visibility_Illustration}(c) shows the effect of coverage for different MAPL values under different possibilities of visibility for the optimum gNB locations. As expected, coverage area monotonically increases with the increase in $\gamma_{\max}$. With direct visibility alone, the fraction of coverage plateaus beyond a threshold of $105$~dB. This is because the coverage area cannot be improved further because of the direct visibility restriction in the pilot problem. A slight increase in the coverage area is possible, via indirect visibility, when there is a favorable scatterer producing specular reflection; this is indicated by the green line. Further, if the scatterers are of diffuse scattering then there is further increase in the coverage area as it can reach more number of points. However, this should be taken with a grain of salt, as these results are based on the same path loss models for both the specular and diffuse scattering regions.  

There are two main benefits of capturing the reflection effects in path loss calculation for the gNB deployment problem. First, exploiting the geometry to identify reflection dominated regions helps better estimate the coverage area. For instance, methods proposed in~\cite{palizban2017automation,szyszkowicz2016automated,mavromatis2019efficient,dong2019cost} would only capture the LoS area of coverage (area within the red boundary) due to their inherent construction which fail to capture the reflection effects. Second, the improved coverage area estimation due to NLoS effects may help to reduce the number of reflectors required to serve the $\mathcal{L}^\text{OSA}$ as it reduces the number of grids in the SAs to be served by the help of reflectors.
 
 \begin{figure}[t!]
    \begin{subfigure}{.5\textwidth}
    \centerline{
        \includegraphics[trim=0cm 0cm 0cm 0.7cm, clip,scale=0.43]{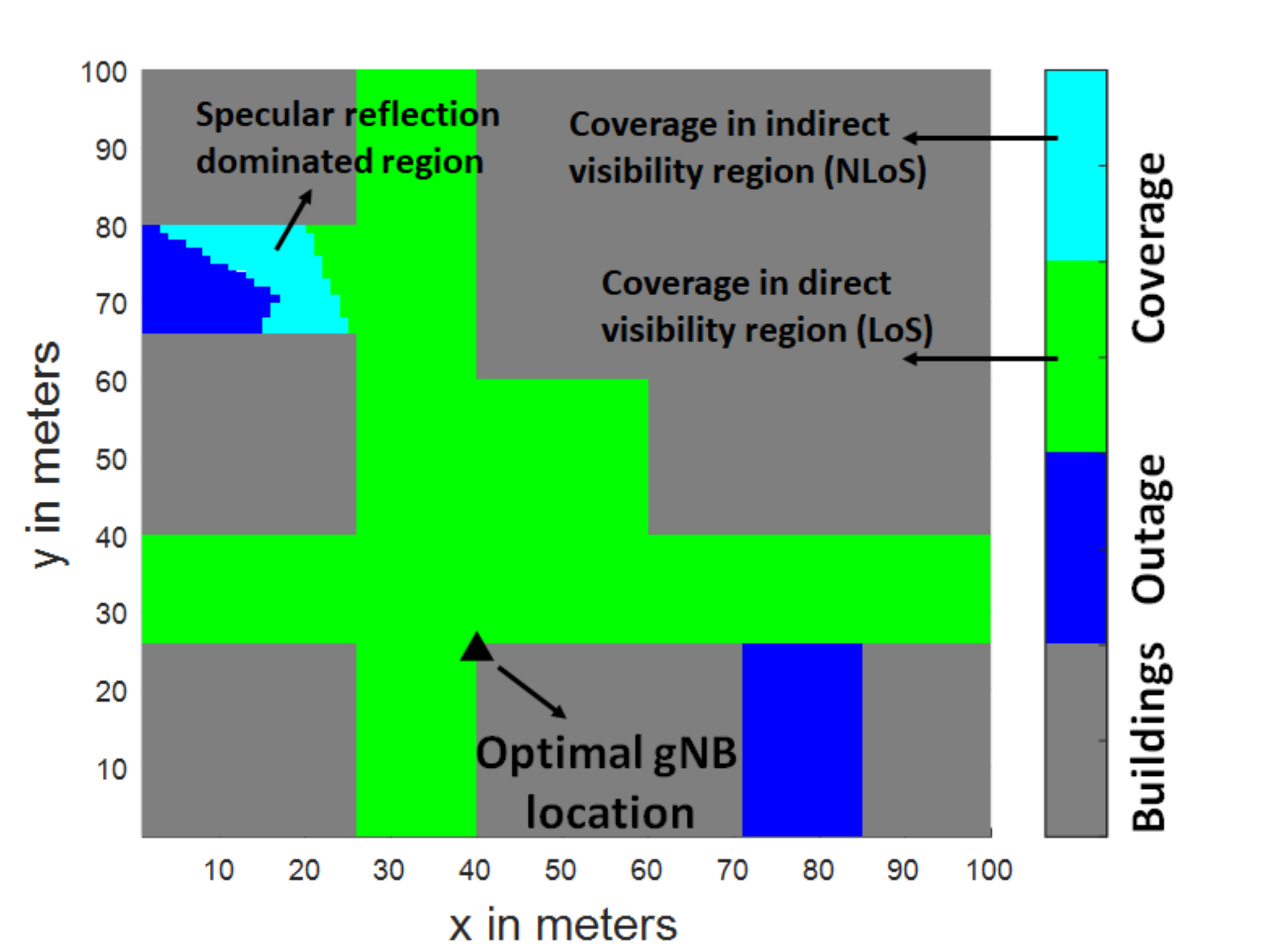}}
            \vspace{-.25cm}
        \caption{ }
    \end{subfigure}
        \begin{subfigure}{.5\textwidth}
    \centerline{\includegraphics[trim=0.1cm 0cm 0cm 0.2cm, clip, scale=0.47]{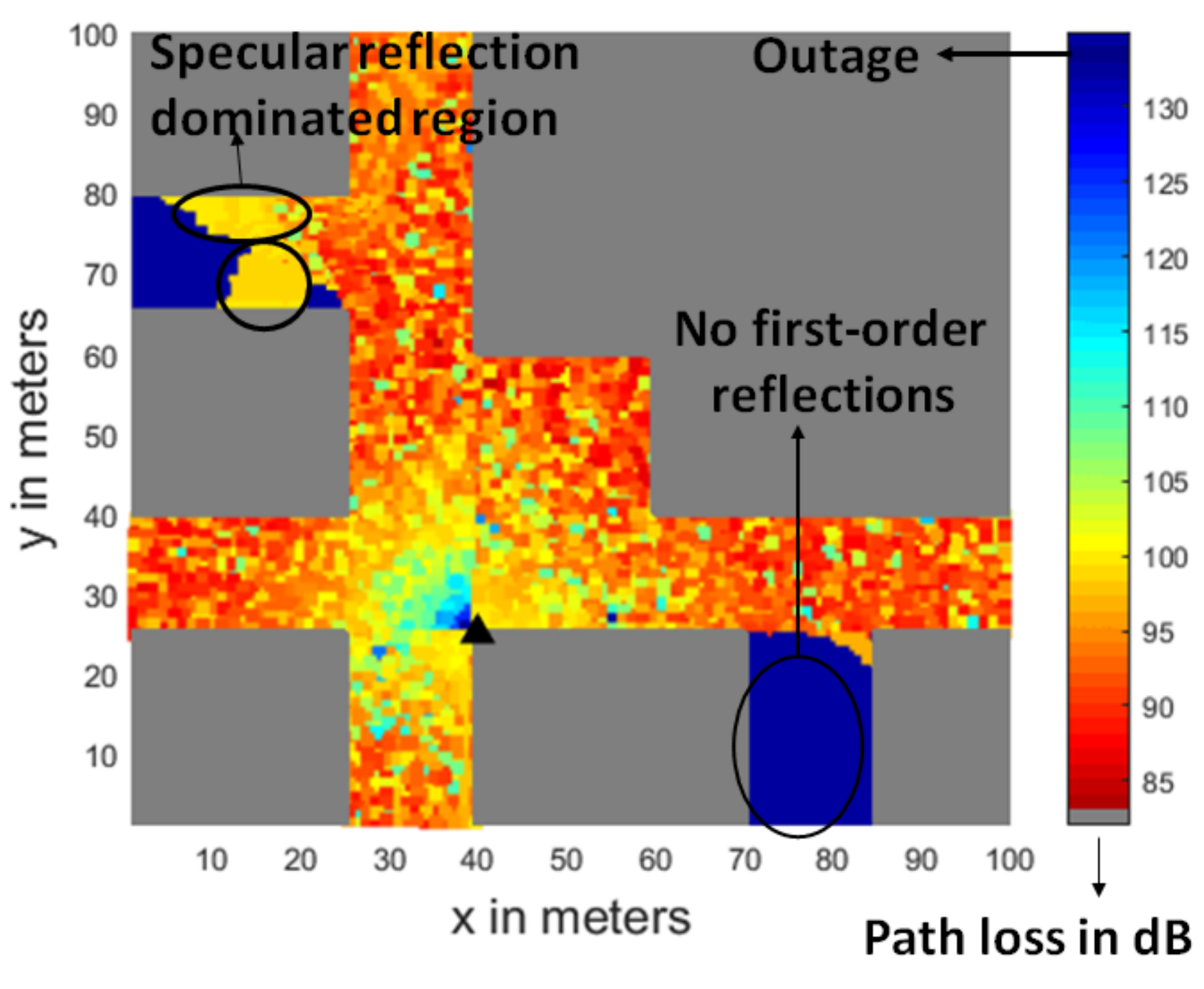}}
    \vspace{-.25cm}
    \caption{}
    \label{fig:Ray_tracing_Comparison}
    \end{subfigure}
    \caption{(a) Coverage map for the fixed gNB location obtained from (\ref{Eq:gNB_Coverage_Problem}) with specular visibility scenario with $\gamma_{\max} = 114$ dB. (b) Average path loss of RT simulation with one reflection. Heatmap from RT simulation.}
    \label{Fig:Compare_RT_vs_Sim}
\end{figure}

\begin{figure*}[t!]
\begin{subfigure}{.33\textwidth}
\includegraphics[scale=0.265]{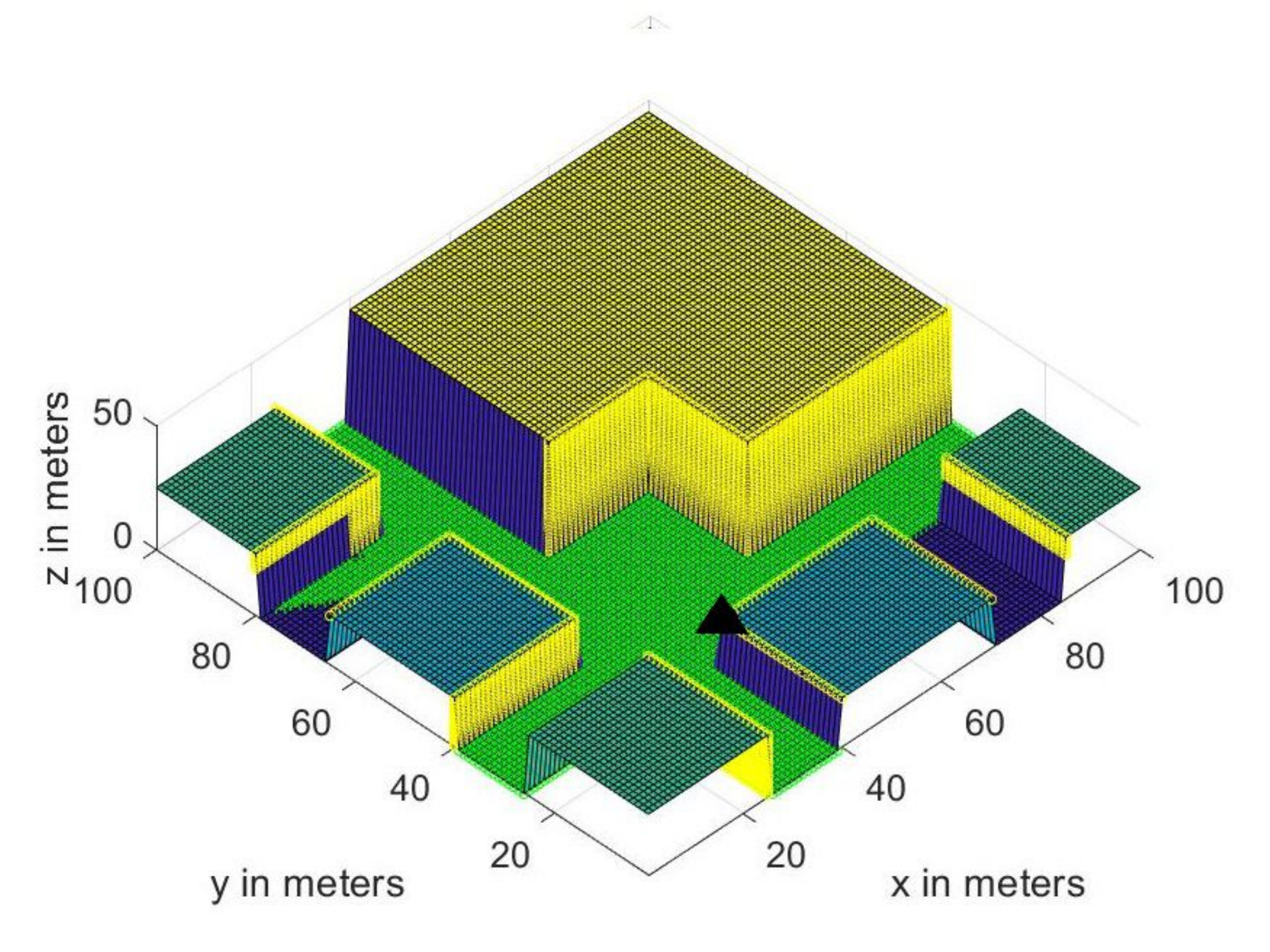}
\caption{ }
\end{subfigure}
\begin{subfigure}{.33\textwidth}
\includegraphics[scale=0.265]{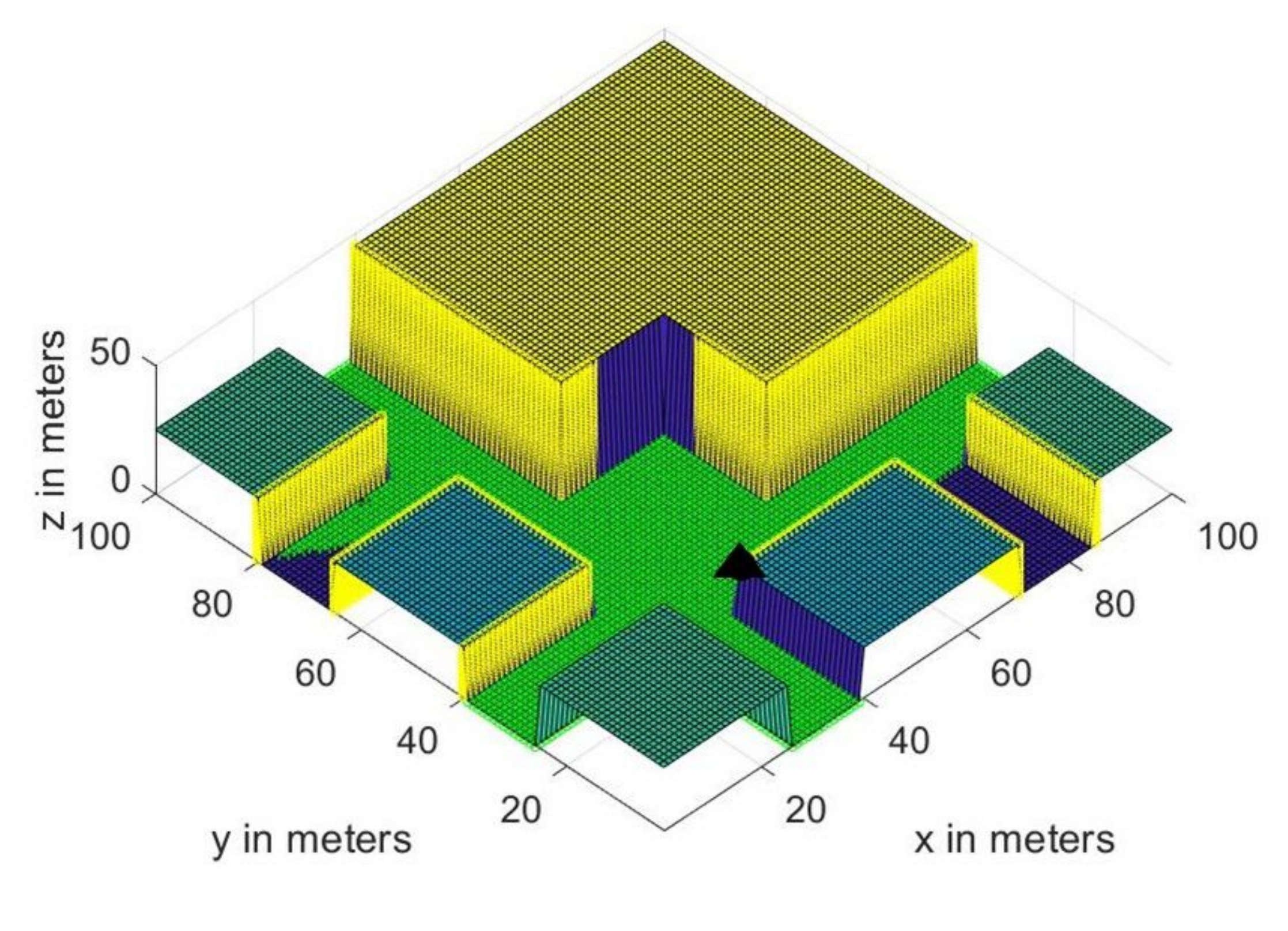}
\caption{ }
\end{subfigure}
\begin{subfigure}{.33\textwidth}
\includegraphics[scale=0.265]{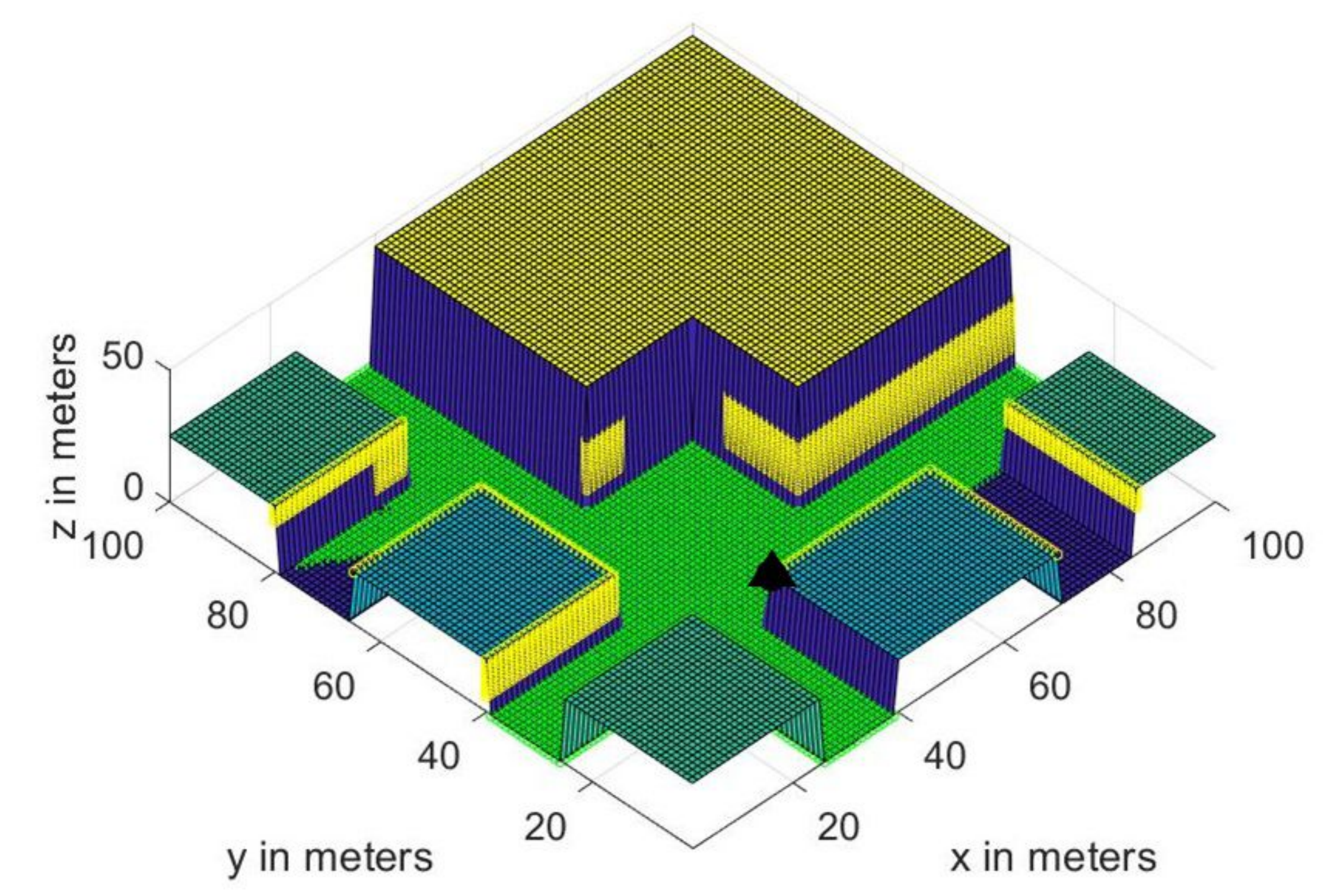}
\caption{ }
\end{subfigure}
\caption{The black triangle corresponds to the optimum gNB location $\mathcal{L}^\text{gNB$_\star$}$. The blue points correspond to the non-coverage points ($\mathcal{L}^{\text{OSA}}$) from the gNB $\mathcal{L}^\text{gNB$_\star$}$. The yellow points correspond to visibility region from (a) $\mathcal{L}^{\text{VgNB$_\star$}} \bigcap \mathcal{L}^\text{Build}$, (b) $\mathcal{L}^{\text{VOSA}} \bigcap \mathcal{L}^\text{Build}$, and (c) correspond to the restricted PMR potential candidates~$\mathcal{L}^\text{PMR}$.}
\label{Fig:BS_Optimization_PMR_candidates}
\end{figure*}
\subsubsection{gNB BILP Optimization}
One such scenario is illustrated in Fig.~\ref{Fig:Compare_RT_vs_Sim}(a). The black triangle corresponds to the choice of the gNB location which provides maximum coverage of the SA under indirect specular reflection with a threshold of $\gamma_{\max} = 114$~dB (guarantees SINR of at least $7$ dB). The grids with green color correspond to the locations that can be covered by the gNB; where the green and cyan boundary indicates the points that are in the direct and indirect (specular) visibility region from the placed gNB. It is evident from the result that considering reflection effects (enabling good reflectors) into path loss calculation helps to capture the area that can be dominated with reflections naturally due to the inherent geometry (without placing reflectors). The blue points correspond to the grids that cannot be covered by the gNB due to specular reflection and are intended to be covered via the~PMRs. \looseness =-1

\subsubsection{Comparison Against RT Solution}
We also created the considered scenario in the RT simulator for the site-specific radio wave propagation analysis. In the past, RT simulators have been widely used for successfully predicting site-specific mmWave propagation, and generated simulations have been used to benchmark against the proposed methods~\cite{khawaja2020coverage,fatih2020MABP}. 

In the RT simulations, we assumed the material of all the scatterers is glass (corresponding to the specular visibility scenario) with dielectric properties of thickness $=3$~mm, permittivity $=6.27$, and conductivity $=2.287 e^{-1}$ defined by the ITU model at $28$ GHz. The number of reflections, transmission, and diffraction were set to $1$, $0$, $0$,~respectively. We deployed the gNB at the location obtained from solving~(\ref{Eq:gNB_Coverage_Problem}). The resulting heatmap from the RT simulations is shown in Fig.~\ref{Fig:Compare_RT_vs_Sim}(b) and used as a benchmark for the proposed geometry-based model. Thus, Fig.~\ref{Fig:Compare_RT_vs_Sim}(a) and Fig.~\ref{Fig:Compare_RT_vs_Sim}(b) can be used for side-by-side comparison.

In Fig.~\ref{Fig:Compare_RT_vs_Sim}(b), the red points on the heat map correspond to the grids, which receive LoS paths indicating direct visibility region from the deployed gNB. Whereas, the specular reflection dominated region is indicated by light orange points (circled points) signifying the region contains significant first-order reflections. This observation is also supported by our experimental result shown in Fig.~\ref{Fig:Visibility_Illustration}(c), which illustrates the same phenomenon, signifying that the proposed path loss model incorporates the knowledge of geometry clearly. Unlike the RT simulation, which calculates the received power based on the detailed characterization of all the MPCs, we propose a solution that captures the geometry and standard 3GPP path loss~models. Thus, our proposed method does not suffer from high computation time and unavailability of detailed and reliable environment description databases, which are essential for the RT simulations.

\subsection{PMRs Deployment}\label{Sec:PMR_Results}
For solving (\ref{Eq:PMR_Optimization}), the size of $a^R$ was set to $0.1$~m (approximately $10 \lambda$). The path loss exponent $\zeta$ was chosen to be equal to $2$ indicating free-space path loss as adopted in~\cite{ozdogan2019intelligent} and verified in~\cite{khawaja2020coverage}.  

\begin{figure}[!t]
\vspace{-.5cm}
    \centerline{\includegraphics[scale=.65]{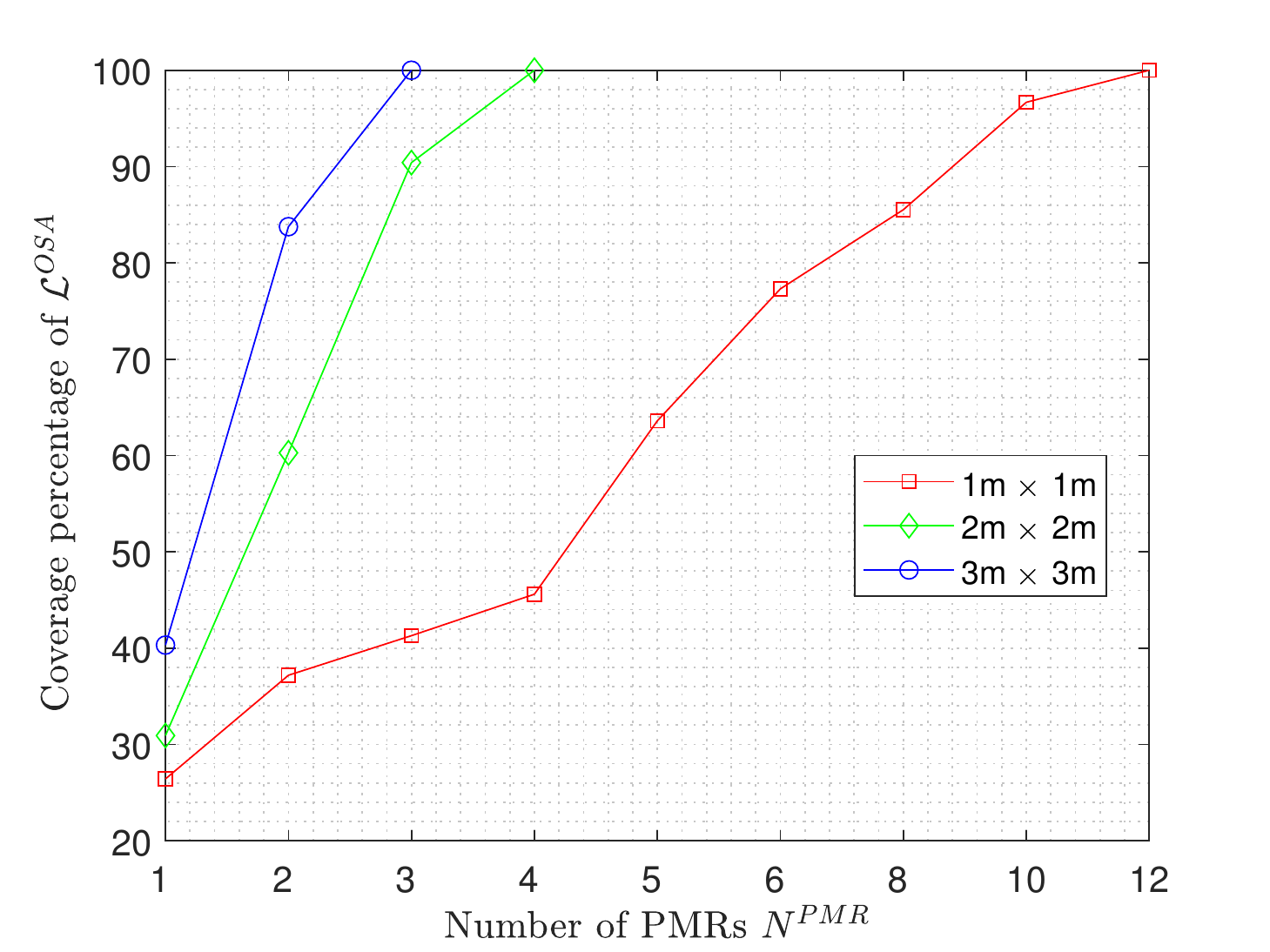}}
    \caption{The fraction of SA in $\mathcal{L}^\text{OSA}$ covered by $N^\text{PMR}$ PMRs of different size with $\gamma_{\max} = 114$~dB.}
    \label{fig:Coverage_Percentage_Refelctors}
\end{figure}

\begin{figure*}[!h]
    \begin{subfigure}{.33\textwidth}
\centerline{\includegraphics[trim=0cm 0cm 0cm 0.9cm, clip,scale=.3]{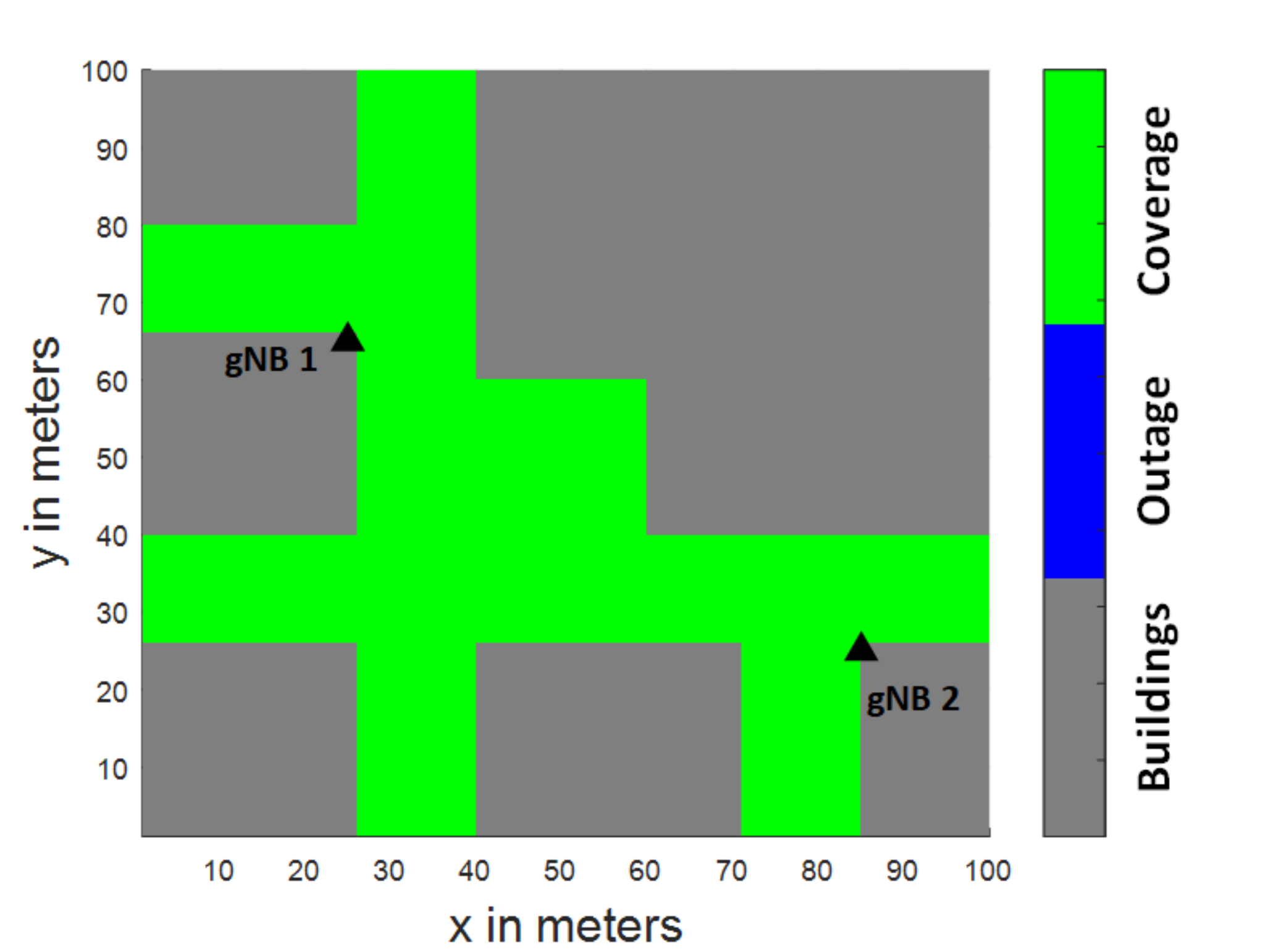}}
    \caption{ }
    \end{subfigure}
    \begin{subfigure}{.33\textwidth}
    \centerline{\includegraphics[trim=0cm 0cm 0cm 0.3cm, clip,scale=.32]{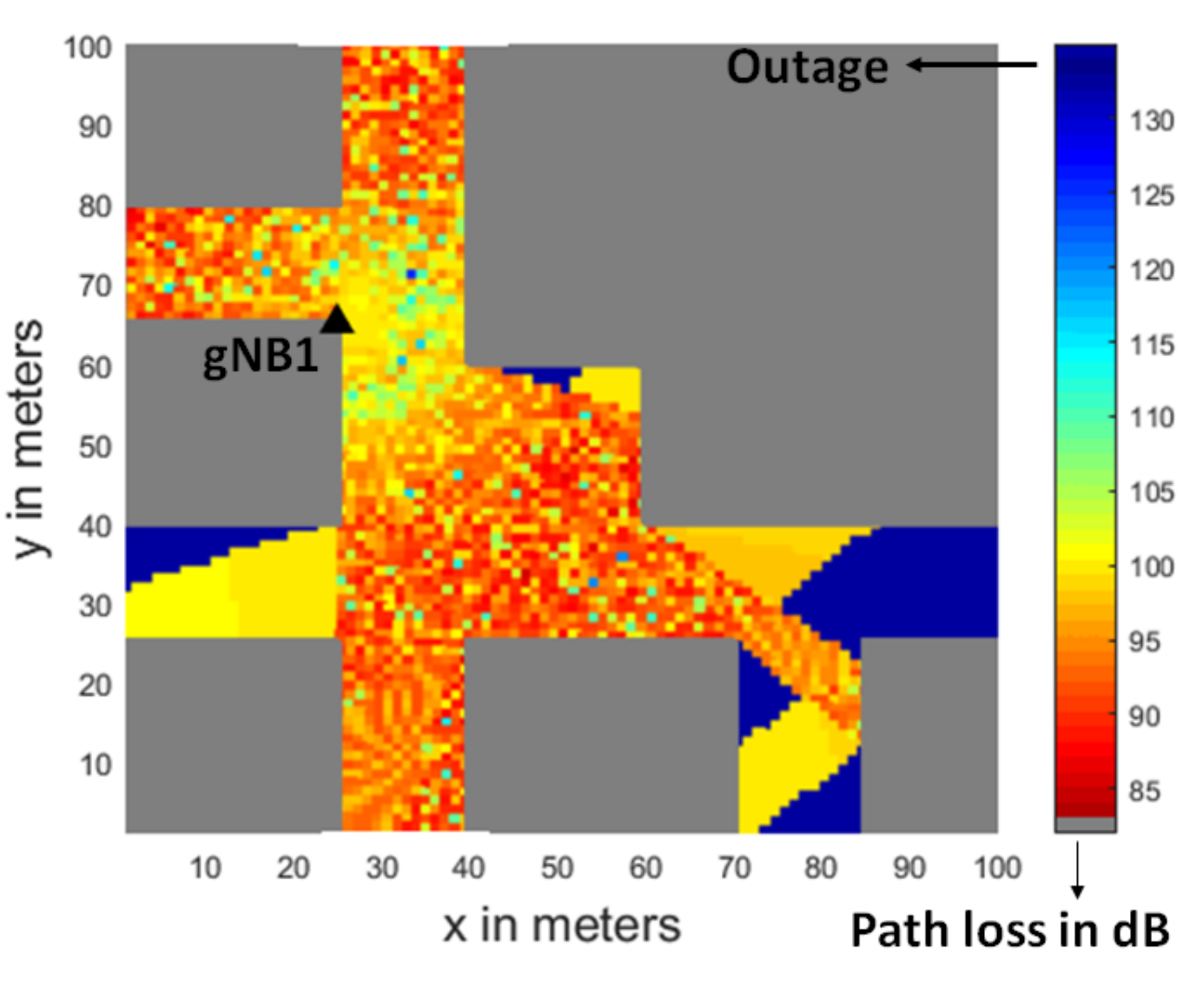}}
     \caption{ }
     \end{subfigure}
         \begin{subfigure}{.33\textwidth}
    \centerline{\includegraphics[trim=0cm 0cm 0cm 0.3cm, clip,scale=.32]{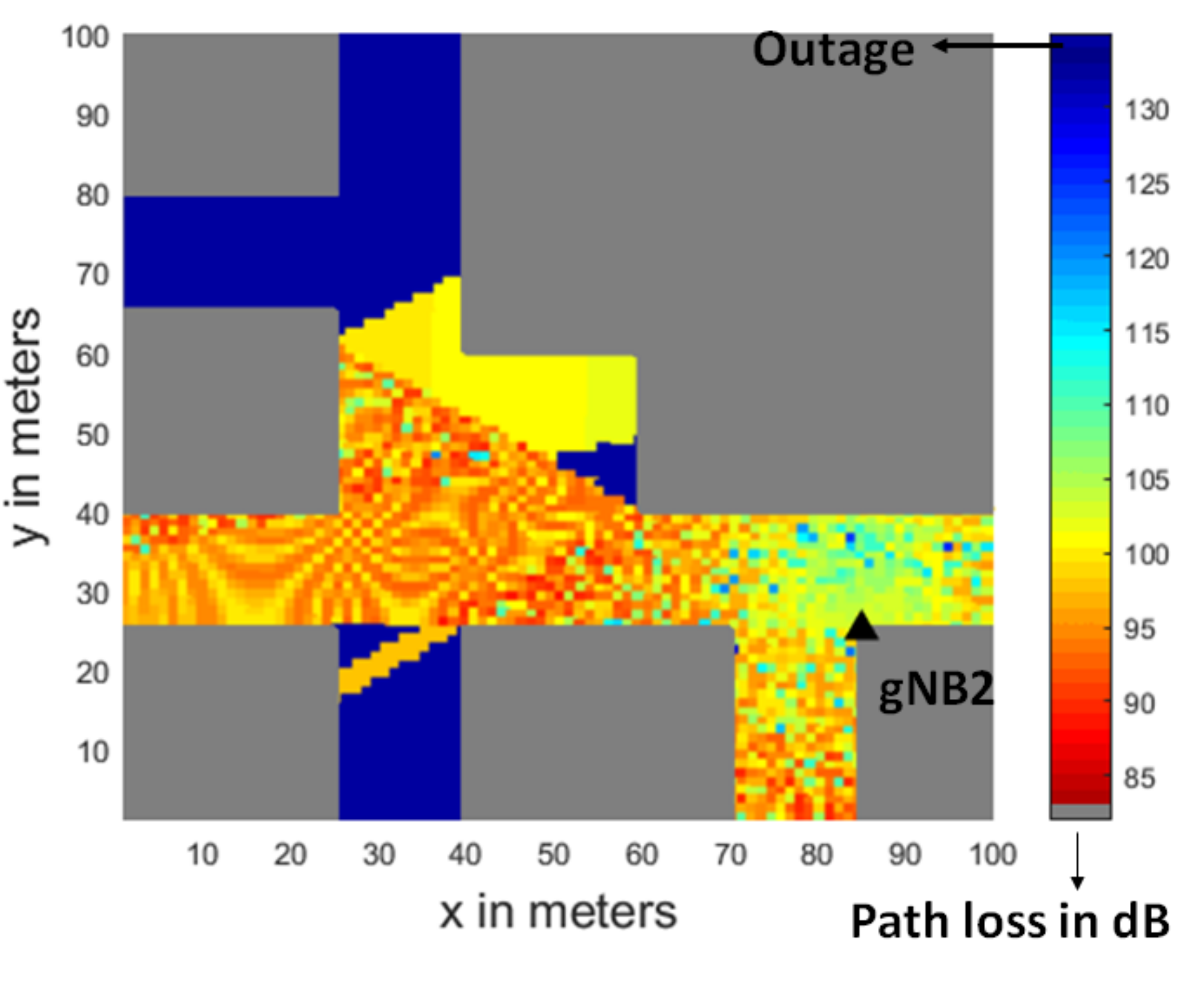}}
     \caption{ }
     \end{subfigure}
    \caption{(a) Coverage area with two gNBs and no reflectors. Heatmap of (b) gNB$_1$ and (c) gNB$_2$ from the RT simulation.}
    \label{fig:2gNB}
    \vspace{-.3cm}
\end{figure*}

\subsubsection{Candidate PMR Locations}
As discussed in Section~\ref{Sec:Potential_PMR_Lcoations}, we restrict the placement of reflectors on the buildings as indicated by the yellow-colored points on the buildings in Fig.~\ref{Fig:BS_Optimization_PMR_candidates}. Fig.~\ref{Fig:BS_Optimization_PMR_candidates}(a) illustrates the points in $\mathcal{L}^\text{gNB$_\star$} \bigcap \mathcal{L}^\text{Build}$, which is the set of all the points that are visible on the buildings from the gNB location. Similarly, Fig.~\ref{Fig:BS_Optimization_PMR_candidates}(b) shows $\mathcal{L}^{\text{VOSA}} \bigcap \mathcal{L}^\text{Build}$, the intersection of the visibility region from the non-covered points and the buildings. The placement of reflectors can also be restricted. For instance, one might desire the reflectors to be above a certain height ($z^\text{PMR,L}$) so that the reflector is visible to the deployed gNB with a low probability of blockage. Also, it is desirable to restrict its placement below a certain height ($z^\text{PMR,H}$) so that the distance between the gNB and SAs through a reflector is not unnecessarily high. Note that by construction $z^\text{PMR,H} > z^\text{PMR,L}$. For the simulation, we set the restricted heights $z^\text{PMR,L}$ and $z^\text{PMR,H}$ of the reflectors to $5$m and $35$m, respectively. The restricted potential candidates of the reflectors $\mathcal{L}^\text{PMR}$ are illustrated in Fig.~\ref{Fig:BS_Optimization_PMR_candidates}(c).

\subsubsection{Effect of Number of PMRs $N^\text{PMR}$ and Size $a^\text{PMR}$}
Next, we show the effect of $N^\text{PMR}$ and square reflector size $a^\text{PMR}$ on the coverage rate of $\mathcal{L}^\text{OSA}$. The coverage rate is obtained after solving (\ref{Eq:PMR_Optimization}) for the required threshold $\gamma_{\max}$. We consider three different values for $a^\text{PMR}$, i.e., $1$~m, $2$~m, and $3$m, and vary $N^\text{PMR}$ until the target QoS threshold and the target coverage area are met. One critical point to take into account while changing~$a^\text{PMR}$ is that the square reflector center point should be~$a^\text{PMR}$ meters away from the buildings due to the orientation requirement (unless it is oriented parallel to the outer surface of the building). This requirement could be a problem for a large-sized PMR (unless it is placed on the boundary of the rooftop). \looseness =-1

Fig.~\ref{fig:Coverage_Percentage_Refelctors} illustrates the effect of~$N^\text{PMR}$ and $a^\text{PMR}$ on coverage rate of $\mathcal{L}^\text{OSA}$ for $\gamma_{\max} = 114$ dB. It is easily seen from Fig.~\ref{fig:Coverage_Percentage_Refelctors} that the increase in the number of reflectors increases the coverage monotonically. This is because, the received power at the considered points increases with $N^\text{PMR}$, which, in turn, helps to meet the QoS constraint. It is also noticed that the higher the dimension of the reflector, the fewer the number of reflectors required to meet the QoS constraint for the considered scenario. Note that the size and number of reflectors may depend on other constraints, such as placement and permission~restrictions.

\subsubsection{Benefits of PMR Placement on Coverage}
For the considered scenario, it is evident from the results that we can cover the entire SA (for $\gamma_{\max} = 114$~dB) with a single gNB and at least $12$ reflectors of size $1$~m (need a lesser number of reflectors if it is of higher dimension). However, it requires at least two gNBs, as shown in Fig.~\ref{fig:2gNB}(a), if the reflectors are not considered. Fig.~\ref{fig:2gNB}(a) is obtained by considering the coverage with direct visibility alone (maximizing LoS area). Fig.~\ref{fig:2gNB}(b)-(c) shows the RT heatmap associated with gNB$_1$ and gNB$_2$, respectively. It can be observed from the heatmaps that the entire coverage area can be covered with $2$ gNBs completely. Thus, considering the reflection effects and PMRs reduces the deployment cost.

\section{Conclusion}\label{Sec:Conclusion}
In this work, we proposed a novel 3D geometry-based framework for the placement of mmWave BSs and PMRs with an objective of enhancing/extending the coverage area with minimum cost. We introduced a GB-PLM that incorporates visibility analysis (direct and indirect visibility) to capture the critical geometry and reflection aspects of the environment. We showed that considering the first-order reflections provides more accurate coverage analysis and hence helps to reduce the number of PMRs required to cover the NLoS area. We also showed that it is possible to achieve the same coverage performance with fewer gNBs by using sufficient number of PMRs at the determined locations. As a result, the proposed solution can significantly reduce capital and operating expenses for mmWave networks.  As a part of the future work, we will look into solving the gNB and PMR placement in a joint manner, with a larger number of gNBs for real-world maps.

\bibliographystyle{IEEEtran}

\bibliography{IEEEabrv,references}

\end{document}